\documentclass[reprint,superscriptaddress,amsmath,amssymb,aps,longbibliography]{revtex4-2}

\usepackage{booktabs}
\usepackage{amsmath}
\usepackage{amssymb}
\usepackage{amsfonts}
\usepackage[caption=false]{subfig}
\usepackage{graphicx}
\usepackage[colorlinks=true, linkcolor=blue, citecolor=blue, urlcolor=blue, plainpages=false]{hyperref}
\usepackage[all]{hypcap}
\usepackage[utf8]{inputenc}
\usepackage{placeins}
\usepackage[english]{babel}
\usepackage[T1]{fontenc}
\usepackage{braket}

\newcommand{\PaperTitle}{Resolving and Suppressing Temporally Correlated Noise in Quantum Error Correction}
\newcommand{\PaperAuthors}{Google Quantum AI and Collaborators}

\renewcommand{\sec}[1]{Section~\ref{sec:#1}}
\newcommand{\fig}[2][]{Fig.~\ref{fig:#2}\textbf{#1}}
\newcommand{\tbl}[1]{Table~\ref{tbl:#1}}

\newcommand{\suppsec}[1]{(see SM sec.~\ref{sec:#1})}

\begin{document}

\title{\PaperTitle}
\date{\today}
\author{\PaperAuthors}

\begin{abstract}
Scaling fault-tolerant quantum error correction (QEC)~\cite{kitaev_fault-tolerant_1997, fowler_surface_2012, terhal_quantum_2015} requires suppressing errors in both space and time, yet standard logical memory experiments primarily probe spatial faults. 
Here, we use a 105-qubit Willow superconducting quantum processor \cite{GoogleQuantumAI2024} to realize and scale the surface code \textit{stability experiment} \cite{Gidney_stability}, which is alternatively sensitive to temporal error chains.
Scaling from patch sizes of $3 \times 3$ to $7 \times 7$, we realize a timelike block suppression factor of $\mu_T = 1.47 \pm 0.25$,
demonstrating the ability to maintain temporal error suppression as the number of qubits increases.
By extending the stability experiment for many cycles, we uncover a sum-of-exponentials behavior in error suppression caused by a temporally correlated error: data-qubit heating to the third excited state $\ket{3}$.
We fully capture this effect with a parameter-free, device-level leakage model incorporating higher-state transport during syndrome extraction cycles.
To suppress this error, we develop and realize a \textit{walking stability circuit}, reaching a logical error probability floor of $\leq 3.20\times 10^{-8}$ at 60 cycles---two orders of magnitude below the standard construction---and limited by ionizing radiation error bursts. 
Finally, by adaptively terminating syndrome extraction based on decoder confidence in an offline analysis, we reduce spacetime logical error by 38\% and time overhead by 34\% compared to the optimal fixed-cycle strategy.
\end{abstract}

\maketitle

\vspace{-6mm}

\setlength{\textfloatsep}{10pt plus 4pt minus 3pt}
\setlength{\floatsep}{4pt plus 2pt minus 2pt}
\setlength{\intextsep}{6pt plus 2pt minus 2pt}

\section{Introduction} \label{sec:intro}

Quantum error correction (QEC) will allow the realization of fault-tolerant quantum algorithms using noisy physical qubits by suppressing logical error rates far below physical error rates \cite{shor_scheme_1995}. 
The standard experimental benchmark in QEC is the so-called memory experiment, in which a logical qubit is initialized, repeated cycles of error correction are performed, and a decoder is used to correct the logical observable measured at the end of the experiment \cite{Krinner2022, Acharya2023, Miao2023, GoogleQuantumAI2024, Eickbusch2025, He2025, Muniz2025, AtomComputing2026, Caune2026, Drouet2026, Xin2026,
  Zhang2026}. 
However, fault-tolerant computation will involve not only qubit preservation, but also controllable logical operations on encoded qubits. 
Recent experiments have focused on realizing and modeling these fault-tolerant logical operations \cite{Postler2022, Bluvstein2024, Reichardt2024, RyanAnderson2024, Daguerre2025, Lacroix2025, Rosenfeld2025, SalesRodriguez2025, Zhou2025, Besedin2026,
  Harper2026, Lin2026, McLauchlan2026, Paetznick2026}, which can often have vastly different sensitivities to physical error mechanisms relative to memory experiments.
It is thus crucial to benchmark, model, and suppress all physical error mechanisms that will limit fault-tolerant operations, as a focus on memory alone can lead to an over-optimization of system design toward logical qubit preservation and is blind to architectural tradeoffs that may improve memory while worsening the fidelity of logical operations. 

Notably, memory experiments are often most sensitive to spatial physical errors that occur across multiple qubits at a similar time---the worst being high-weight errors spanning multiple qubits and flipping an observable.
Logical operations, on the other hand, can have regions in spacetime that are instead most sensitive to temporal errors that occur at a similar spatial location across many cycles of error correction.
Benchmarking device performance using simple experiments that probe logical sensitivity to both spatial and temporal errors would allow for a window into both of these mechanisms.
Importantly, uncovering any low-probability correlated errors now (an example being high energy events \cite{mcewen_resolving_2022, Kurilovich2026, Acharya2023,GoogleQuantumAI2024}) will allow for hardware mitigation before scaling up. 

\begin{figure}[t!]
    \centering
    \includegraphics[width=\columnwidth]{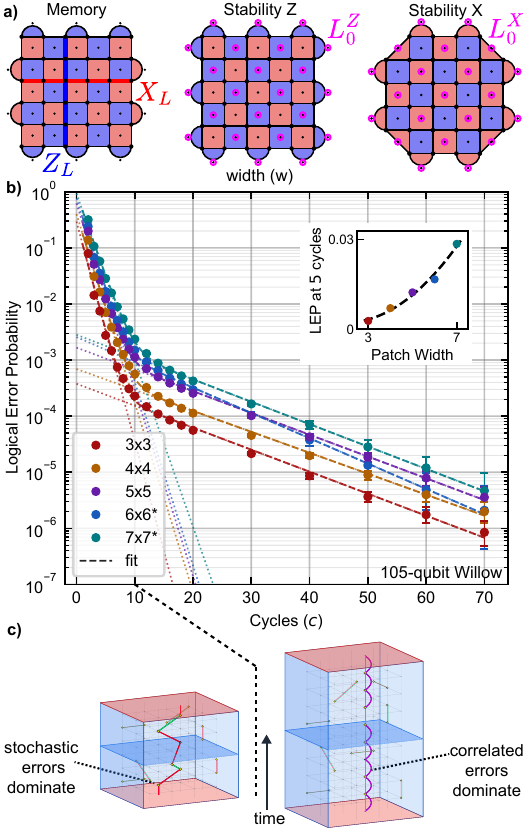}
    \caption{
    \textbf{a)} Memory and stability surface code patches. Measure (data) qubits are small (large) dots. $X$-type ($Z$-type) stabilizers are colored red (blue). The $X$ patch has cut corners in order to maintain the checkerboard in the bulk, which sets the gate ordering.
    \textbf{b)} Measured stability logical error probability on 105-qubit Willow with correlated matching decoding.  Data are averaged across 7, 4, 2, 1, and 1 subgrids for patch width $w = 3, 4, 5, 6,$ and $7$, respectively, with each subgrid and basis running a 200-cycle stability circuit for $2 \times 10^{6}$ raw shots. Extracting five subsampling windows per shot yields $10^{7}$ ($4.5 \times 10^{6}$) effective shots per circuit for $c \le 20$ ($c = 70$). 
    *Due to device constraints, $6 \times 6$ and $7 \times 7$ grids are $Z$-basis stability experiments only; all other grids are averaged over $Z$ and $X$ bases.
    \textbf{Inset}: quadratic scaling of logical error with patch width after 5 cycles. \textbf{c)} Illustration of Pauli and correlated error regimes.} 
    \label{fig:1}
\end{figure}

The stability experiment \cite{Gidney_stability} is the spacetime dual to the memory experiment and is most useful for benchmarking the impact of temporal errors on logical operators.
Importantly, it can be used together with memory experiments to predict the performance of primitive logical operations such as lattice surgery and logical qubit moves \cite{Domokos2024}. 
As such, the stability experiment is another crucial benchmark that should be run on all devices and should be treated on equal footing with the memory experiment for scaling up logical operations.
Recent experiments have begun exploring stability and its repetition-code-like variants \cite{Caune2026, Xin2026, Drouet2026, Harper2026, McLauchlan2026}.

In this work, we scale surface code stability experiments up to a full-device $7\times 7$ patch and hundreds of error-correction cycles on Google's 105-qubit Willow processor.
Using the stability experiment alongside memory experiments, we introduce and benchmark the \textit{spacetime error}, which is sensitive to both temporal and spatial errors. 
Using device-informed models, we reveal the relative importance of various physical error channels between spacelike and timelike observables, including the relatively high importance of readout and reset in the stability experiment.
Additionally, we use the stability experiment as a metrology tool, uncovering low-probability temporally correlated errors.
We then fix these errors in experiment using a walking stability circuit~\cite{mcewen_relaxing_2023, Eickbusch2025}, 
realizing a minimum logical error probability of $\leq 3.20\times 10^{-8}$ at 60 cycles---suppressing logical error by two orders of magnitude compared to the standard circuit compilation at 60 cycles. Finally, we use the decoder to analyze dynamic stopping strategies \emph{ex post facto}. Using soft information from the decoder \cite{gidney_yoked_2023, akahoshi2026runtime}, we demonstrate a $38\%$ reduction in error and a $34\%$ reduction in mean runtime when minimizing error.

\vspace{-2mm}
\section{Stability experiment regimes} 
\label{sec:stability}

To understand the stability experiment, we first compare it to the more well-known memory experiment.
A standard patch of surface code encoding a distance $d=6$ logical qubit for a memory experiment is shown on the left of \fig[a]{1}. 
In this patch, data qubits and measure qubits are aligned in a square grid, with the top and bottom (left and right) boundary weight-2 stabilizers being $Z$-type ($X$-type).
In such a configuration, the logical operators $X_L$ ($Z_L$) are defined as products of data-qubit $X$ ($Z$) operators connecting the left and right (top and bottom) edges of the patch \cite{fowler_surface_2012}. 
In contrast to the memory experiment, the stability experiment's patch of surface code does not encode a logical qubit, as shown in the right two grids in \fig[a]{1}.
Instead, stability has only one type of boundary stabilizer defining the basis $b$ of the experiment.
In such a configuration, the $b$-basis stabilizers $(S)$ are overdetermined, and their product satisfies the global constraint $L_0 = \prod_{i}S_i = 1$, defining a logical observable $L_0$.
The stabilizers of these patches are repeatedly measured for $c$ cycles, and the results are fed to a decoder to correct errors in the measurement of $L_0$, which can be defined at any arbitrary cycle.
Repeatedly measuring the stabilizers for $c$ cycles defines a decoding graph for which $\lfloor(c - 1)/2\rfloor$ errors across time can be corrected, and hence the stability experiment has a fault distance $d=c$.

To benchmark the performance of stability in hardware, we run stability experiments with data-qubit width $\times$ height from $3\times3$ to $7\times7$ on a 105-qubit superconducting Willow processor \cite{GoogleQuantumAI2024}. 
Similar to memory scaling experiments, we average the logical results of multiple stability patches of each size across the device to accurately compare the results of low-patch-width experiments (e.g., $3\times3$) to full-device experiments ($7\times7$).
We use the standard 4-layer stabilizer measurement circuit compiled to CZ operations and idle between cycles in the ZXXZ code to remove unnecessary single-qubit gates \cite{GoogleQuantumAI2024}.
We also include dynamical decoupling, measure-qubit leakage removal using multi-level reset (MLR) \cite{McEwen2021}, and data-qubit leakage removal (DQLR) of the $\ket{2}$ state using leakage-iSWAP and fast measure-qubit reset operations \cite{Miao2023}.

The measured results of our stability-scaling experiment across patch sizes, decoded with a correlated matching decoder \cite{higgott_sparse_2023} and an SI1000 error model \cite{Gidney_2021_honeycomb} prior, are shown in \fig[b]{1}, where the logical error probability is averaged over multiple patches of the same size and across bases ($Z$ and $X$)\footnote{Except for $w=6, 7$, for which device layout constraints limit the experiment to the $Z$ basis only.} (see SM secs.~\ref{sec:stability_supplement} and~\ref{sec:extended_data_xz_grids}).
To increase the effective number of shots, we employ a \textit{subsampling} technique in which short-duration stability-experiment data is taken as multiple windows from longer stability experiments (200 cycles for the scaling dataset and 1000 cycles for the walking experiment) \suppsec{supp_stability_subsampling}.
We calibrate our experiment using reinforcement learning \cite{sivak2026reinforcement} with a cost function that includes both memory and stability detection probabilities \suppsec{supp_rl_calibration}.

Two prominent features are shown in the data.
First, unlike in a memory experiment, the measured stability logical error probability at a fixed number of cycles worsens as patch size increases in space, as shown in the inset of \fig[b]{1}. 
This is expected from a simple path-counting argument; the number of possible error chains that can cause a logical error increases approximately linearly with the number of qubits, and the resulting logical error probability at a fixed number of cycles increases quadratically with respect to the patch width.

Second, our data differs from a single exponential suppression with cycles, which is expected for uncorrelated local errors given the distance increase with cycles.
After around $10$ cycles, the logarithm of the logical error probability, $\log\left(\text{LEP}\right)$, incurs a \textit{knee}, after which the suppression worsens. 
To quantify performance in both regimes, we fit our data to a sum-of-exponentials model, $P_L(c) = a_1 s_1^{-c} + a_2 s_2^{-c}$, as motivated under a simple correlated error model \suppsec{correlated_stability_model}, with results given in \tbl{1}. 
Notably, the exponential suppression after the knee, $s_2\approx 1.1$, is consistent across patch widths, indicating the same mechanism is likely limiting suppression after the knee for all patches.
As will be shown through both detailed noise modeling and direct measurements in \sec{modeling}, this sum-of-exponentials behavior is caused by data-qubit leakage to the third excited state $\ket{3}$, which dominates logical errors after the knee.

The stability experiment and its sum-of-exponentials behavior can be understood in terms of metrology with two regimes.
The Pauli regime (before the knee) is defined as the section where logical error can be accurately predicted with a stochastic Pauli model. The correlated error regime (after the knee) is dominated by low-probability non-Markovian error mechanisms.
Potential sources of these errors include higher-state leakage, measurement-induced state transitions, and coupling to coherent two-level system (TLS) defects. 
Under some correlated error models, the logical error can exhibit multiple knees, with regions where the logical error worsens with cycles before being further suppressed \suppsec{supp_non_markovian_models}.

\begin{table}[t]
\centering
\footnotesize
\setlength{\tabcolsep}{2.8pt}
\begin{tabular}{ccccc}
\hline\hline
Patch Size & $a_1$ & $s_1$ & $a_2\ (\times 10^{-3})$ & $s_2$ \\
\hline
$3\times 3$ & $0.29 \pm 0.07$ & $2.46 \pm 0.12$ & $0.38 \pm 0.05$ & $1.094 \pm 0.004$ \\
$4\times 4$ & $0.40 \pm 0.08$ & $2.17 \pm 0.08$ & $0.69 \pm 0.08$ & $1.090 \pm 0.003$ \\
$5\times 5$ & $0.62 \pm 0.10$ & $2.15 \pm 0.07$ & $1.67 \pm 0.16$ & $1.094 \pm 0.003$ \\
$6\times 6^*$ & $0.77 \pm 0.11$ & $2.13 \pm 0.06$ & $2.58 \pm 0.23$ & $1.109 \pm 0.003$ \\
$7\times 7^*$ & $0.93 \pm 0.10$ & $1.98 \pm 0.04$ & $2.83 \pm 0.21$ & $1.096 \pm 0.002$ \\
\hline\hline
\end{tabular}
\caption{Fit parameters for the two-component stability decay model $P_L(c) = a_1 s_1^{-c} + a_2 s_2^{-c}$ across patch sizes. 
Uncertainties represent $1\sigma$ standard errors.}
\label{tbl:1}
\end{table}

\section{Behavior in the Pauli regime}

Near-term experiments involving logical gates will require a detailed understanding of both the spacelike and timelike error mechanisms in the regime where cycles $\sim d$. 
Hence, we benchmark Willow in this regime using both stability and memory experiments (see \fig[a]{2}) on multiple patch sizes tiling the device.
Here, correlated errors are subdominant, as shown by our stability data in \fig[b]{1}; as a result, a stochastic Pauli model accurately predicts device performance in this so-called Pauli regime.
We show this by constructing a circuit-level stochastic Pauli noise model directly from per-qubit and per-pair device benchmarks~\cite{Eickbusch2025} \suppsec{supp_component_error_rates}. 
Coherent errors and single-cycle leakage effects are Pauli-twirled into effective depolarizing and bit-flip channels before entering the model. 
Simulating this Pauli model across both $X$ and $Z$ bases reproduces the measured logical error probabilities for both memory ($d=3\text{--}7$) and stability ($w=3\text{--}7$) at cycle counts comparable to the code distance (\fig[b]{2}), including the characteristic even--odd cycle staircase in short stability experiments. 
However, the stability data diverges sharply upward from the Pauli prediction after $c = (6, 7, 8, 9, 9)$ cycles for patch widths $w = (3, 4, 5, 6, 7)$, respectively. 
This knee occurs when the cumulative probability of a multi-cycle correlated error event exceeds the exponentially suppressed stochastic Pauli logical error probability. 

\begin{figure*}[t!]
    \centering
    \includegraphics[width=\textwidth]{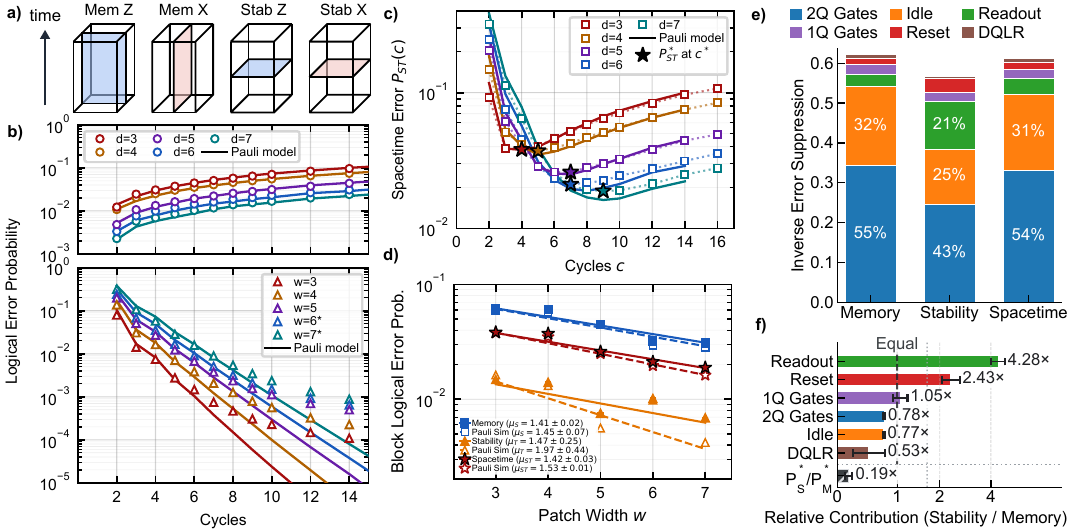}
    \caption{ \textbf{a)} Memory and stability experiment topological diagrams. The colored red and blue sheets represent $X$ and $Z$ logical observables.
    \textbf{b)} Measured memory and stability experiments in the Pauli regime, compared to a Pauli model simulation. 
    \textbf{c)} Measured and Pauli-modeled spacetime error probabilities, with minima $P_{\text{ST}}^*$ at the optimal cycle count $c^*$ shown as stars.
    \textbf{d)} Scaling of the block logical error probabilities ($P^*_M$, $P^*_S$, and $P^*_{\text{ST}}$) evaluated on a $w\times w \times c^*$ spacetime block with patch width $w$ ($=d$ for memory).
    \textbf{e)} Inverse block error suppression ($1/\mu$) budgets for memory, stability, and spacetime under a Pauli model.
    \textbf{f)} Comparison of the relative contribution of various error channels between stability and memory, along with the ratio between stability and memory logical error probabilities at the optimal cycle count ($P^*_S / P^*_M$ at $P_{\text{ST}}^*$).
    }
    \label{fig:2}
\end{figure*}

To quantify the performance relevant for logical operations such as lattice surgery and logical qubit moves, we define the \textit{spacetime error probability} $P_\text{ST}$. 
For both memory ($M$) and stability ($S$) on the same patch of qubits, we take the sum of the logical error probabilities across the $X$ and $Z$ bases, $P_M(c) = P_{M,X}(c) + P_{M,Z}(c)$ and $P_S(c) = P_{S,X}(c) + P_{S,Z}(c)$; the spacetime error is then defined as the average of these sums, $P_\text{ST}(c) = \frac{1}{2}\left[P_M(c) + P_S(c)\right]$. 
For our data, $P_\text{ST}$ is plotted in \fig[c]{2}. 
Since the stability error $P_S(c)$ is exponentially suppressed with cycles, and the memory error $P_M(c)$ increases linearly with cycles to first order, a minimum in $P_{\text{ST}}$ occurs at an optimal cycle count $c^*$. 
To avoid statistical uncertainty when identifying $c^*$ from single-cycle data points, we determine $c^*$ by finding the minimum of the fitted spacetime curve $P_{\text{ST}}(c)$ obtained from the logical error rate fit to $P_M(c)$ and the two-exponential decay fit to $P_S(c)$. 
We call the resulting minimum the \textit{minimum spacetime error} $P^*_\text{ST} = \frac{1}{2}\left(P^*_M + P^*_S\right)$, where $P^*_M \equiv P_M(c^*)$ and $P^*_S \equiv P_S(c^*)$ denote the constituent memory and stability block error probabilities evaluated at the optimum $c^*$.
We find the optimal number of cycles for patch widths $(3,4,5,6,7)$ to be $(4,5,7,7,9)$, indicating that logical operations such as lattice surgery and qubit moves will likely benefit from waiting $\sim d + 1$ or $d+2$ cycles, which is longer than the typical rule of thumb $c^*=d$, and consistent with the balance between timelike and spacelike error mechanisms presented in \cite{Domokos2024}.

Under stochastic Pauli noise, the minimum spacetime error is expected to be exponentially suppressed with increasing patch distance $d$ ($=w$). 
As such, we fit the minimum spacetime error using the suppression
model $P^*_\text{ST} = A\left(\mu_{\text{ST}}\right)^{-(d+1)/2}$, and find a spacetime suppression factor of $\mu_{\text{ST}} = 1.42 \pm 0.03$, 
indicating Willow can suppress the error rate of logical operations that are sensitive to both spatial and temporal errors as the number of qubits is increased.
Only odd-distance points are included in the fit, and the symbol $\mu$ denotes a block suppression factor rather than a per-cycle suppression factor.
We expect better performance can be achieved using the same data under more sophisticated decoding techniques, including ML decoding \cite{Bausch2024learning}.

The measured spacetime suppression is slightly lower than the Pauli-model prediction of $\mu_{\text{ST (Pauli model)}} = 1.53 \pm 0.01$.
This difference is due to error mechanisms such as leakage, which are not captured by the Pauli model. 
Accordingly, we introduce a full leakage model in the next section.
In \fig[d]{2}, we compare the spacetime error suppression to the spacelike block suppression $\mu_S = 1.41 \pm 0.02$ ($\mu_{S\text{ (Pauli model)}} = 1.45 \pm 0.07$) and timelike block suppression $\mu_T = 1.47 \pm 0.25$ ($\mu_{T\text{ (Pauli model)}} = 1.97 \pm 0.44$). 
All block error rates in \fig[d]{2} ($P^*_M$, $P^*_S$, and $P^*_{\text{ST}}$) are evaluated at the same optimal cycle count $c^*$, where a logical block is defined by a spacetime volume of $w \times w \times c^*$, with $w = d$ for the memory experiments. 

Finally, to quantify how individual physical error mechanisms limit performance in the Pauli regime, we construct error budgets for the spatial ($1/\mu_S$), temporal ($1/\mu_T$), and combined spacetime ($1/\mu_{\text{ST}}$) inverse block error suppression factors.
All such error budgets are evaluated at the fixed experimental optimal cycle counts $c^*$ via one-at-a-time channel ablation in our Pauli model (\fig[e]{2}). 
By ablating each error channel family in turn, we partition each $1/\mu$ into contributions from two-qubit gates (2Q Gates), data-qubit idling during measurement and reset (Idle), mid-circuit readout, reset, single-qubit gates (1Q Gates), and data-qubit idle errors during leakage removal operations (DQLR).
While two-qubit gates ($55\%$ in memory, $43\%$ in stability) and data-qubit idling ($32\%$ and $25\%$) represent the largest absolute contributors to both metrics, their relative contributions differ markedly (\fig[f]{2}). 
Temporal chains of measure-qubit flips directly penetrate the stability observable sheet $L_0$; hence, predominantly timelike error channels carry increased fractional weight in stability when compared to memory---specifically reset ($6.0\%$ vs.\ $2.5\%$, a $2.43\times$ relative contribution) and readout ($21.0\%$ vs.\ $4.9\%$, $4.28\times$). 
Conversely, predominantly spacelike data-qubit and other error channels carry reduced or roughly equal relative weight in stability (`2Q Gates' at $0.78\times$, `Idle' at $0.77\times$, `DQLR' at $0.53\times$, and single-qubit gates at $1.05\times$). 

At the minimum spacetime error $P_{\text{ST}}^*$, the optimal cycle count $c^*$ occurs where the steep exponential decay of stability balances the gentler linear growth of memory with cycles; as a result, the stability error $P^*_S$ is substantially smaller than the memory error $P^*_M$ at the optimum, as seen directly in the block error rates of \fig[d]{2}. 
Specifically, stability ($P^*_S$) accounts for $16\%$ of the combined logical error across odd distances compared to $84\%$ from memory ($P^*_M$, corresponding to a $P^*_S / P^*_M \approx 0.19\times$ ratio in \fig[f]{2}); consequently, in the Pauli regime, the combined spacetime error budget ($\mu_{\text{ST}} = 1.53$) closely mirrors the memory budget ($54\%$ two-qubit gates, $31\%$ idling, $6\%$ readout, $4\%$ single-qubit gates, $3\%$ reset, and $2\%$ DQLR), while remaining first-order sensitive to any degradation in temporal error suppression.
The associated $1/\mu$ budgets and single-channel impact on logical error probability for memory, stability, and spacetime are detailed in SM sec.~\ref{sec:supp_error_budgeting}.

\section{Behavior and modeling in the correlated error regime}
\label{sec:modeling}

Although near-term logical performance on smaller surface code patches ($d \lesssim 7$) is governed by Pauli errors, time-correlated errors will limit performance at larger cycle counts, and thus larger logical patch sizes.
In order to build intuition for how such error mechanisms impact the performance of the stability experiment, we consider a minimal phenomenological model in which a local defect occurs with probability $p_{\text{corr}}$ per qubit per cycle and persists for a characteristic lifetime $\tau$ cycles.
The defect, while active, will randomize adjacent stabilizer measurements \suppsec{correlated_stability_model}; a realization that persists for $\ge \lfloor(c+1)/2\rfloor$ cycles is sufficient to cause a logical fault.
This is observed as the stability logical error probability transitions from steep Pauli suppression to a shallow correlated regime governed by the tail of the defect lifetime distribution. 
In this two-regime sum model, the logical error at which the knee appears scales quadratically with patch width and linearly with the defect occurrence rate ($\propto w^2 p_{\text{corr}}$), whereas the post-knee slope is set by the persistence lifetime $\tau$, yielding a severely degraded post-knee suppression factor ($s_2 \approx 1.1$ across $w=3\text{--}7$ in \fig[b]{1}). 
Notably, if the defect duration is concentrated near a fixed length $L$, the stability logical error probability can  increase non-monotonically with cycle count for $c \lesssim 2L$ before decaying.

\begin{figure*}[t!]
    \centering
    \includegraphics[width=\textwidth]{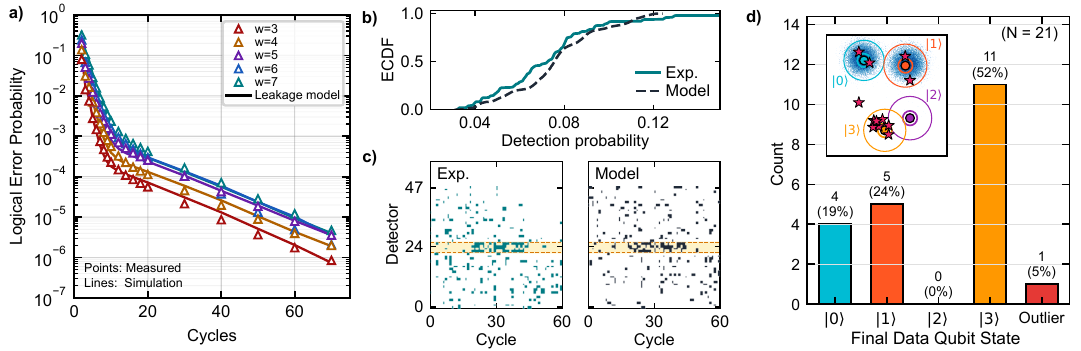}
    \caption{\textbf{Stability in the Correlated Error Regime.}
    \textbf{a)} Leakage model prediction (lines) for stability logical error probability vs.\ cycles, compared to experimental data (triangles).
    \textbf{b)} Cumulative distribution of per-detector firing probabilities across a $7\times 7$ ($c=50$) stability experiment (solid teal, median $0.0726$) and the multi-level simulation (dashed black, median $0.0756$).
    \textbf{c)} Representative individual shots of detection patterns from experiment (left) and simulation (right), showing multi-cycle syndrome bursts across the four measure qubits bordering a $\ket{3}$-leaked data qubit (shaded yellow band).
    \textbf{d)} Final data-qubit state after 14 cycles in the $3\times3$ stability experiment, postselected on limiting correlated error chains.}
    \label{fig:3}
\end{figure*}

While this simple model suffices for capturing the high-level intuition, mitigating such errors on the physical device requires elucidating their underlying physical mechanism.
Although Data Qubit Leakage Removal (DQLR) implemented on Willow processors~\cite{Miao2023} resets the $\ket{2}$ population every cycle with high efficacy, it is not explicitly designed to remove the $\ket{3}$ population \suppsec{supp_numerical_model}.
Consistent with this, we show that the dominant time-correlated error is the low residual $\ket{3}$ population, produced primarily by passive heating within the device.
Heating rates were characterized on the device and included directly within our model; a detailed description and measured rates can be found in SM sec.~\ref{sec:supp_numerical_model}. 
Phenomenologically, leakage to higher excited states behaves similarly to the defect described above.
Once in $\ket{3}$, a data qubit will corrupt adjacent stabilizer measurements across consecutive cycles until it exits the $\ket{3}$ manifold. 
We note that aside from intrinsic relaxation, the dominant $\ket{3}$ removal mechanism is leakage transport during diabatic CZ gates, in which leakage swaps onto the measure qubit, where it is cleared by multi-level reset \suppsec{supp_numerical_model}. 

Incorporating all measured gate component benchmarks and heating rates alongside analytical leakage transport into a single model reproduces the measured stability logical error probabilities across all patch widths ($w=3\text{--}7$) and out to $70$ cycles without any free fitting parameters (\fig[a]{3}). 
Beyond aggregate logical performance, the multi-level model reproduces both the ensemble distribution and single-shot spatiotemporal structure of syndrome detections (\fig[b,c]{3}).
Across all detectors in the $7\times 7$ patch, the simulated detection probabilities match the measured distribution with a median of $7.56\%$ compared to $7.26\%$ in experiment (\fig[b]{3}).
At the single-shot level, the simulated realizations exhibit the same qualitative signature as experimental shots near the error floor: isolated background detections punctuated by abrupt, multi-cycle bursts of high detection density across up to four measure
qubits adjacent to a single $\ket{3}$-leaked data qubit (\fig[c]{3}).
These correlated syndrome patterns persist for $\sim 10\text{--}15$ cycles---consistent with the effective $\ket{3}$ residence time $\tau_{\ket{3}}$ set by the action of intrinsic relaxation and leakage transport.

For primitive logical operations such as lattice surgery and qubit movement requiring $c \approx d$ cycles on a patch of width $w = d$, experiments at $d \le 7$ operate in or near the Pauli regime ($c^* \le 9$), where stability ($P^*_S$) contributes only $16\%$ of the minimum spacetime error $P_{\text{ST}}^*$. 
However, as shown by the model, once the code distance increases past $d \approx 9$, a logical operation of duration $c = d$ crosses the stability knee into the $\ket{3}$-dominated correlated error regime. 
We note that despite spacelike memory errors decreasing rapidly with distance (the memory logical error rate $\varepsilon$ follows $\varepsilon = a\left(\Lambda_S\right)^{-(d+1)/2}$ with suppression factor $\Lambda_S = 2.17$ in the Pauli model and $1.98$ in experiment), we expect post-knee stability errors to grow quadratically with patch width ($w^2 = d^2$) and decay slowly with cycle count.
In short, the correlated timelike error is expected to overtake the memory error at $d^* \approx 11\text{--}13$ ($P_L \sim 10^{-3}\text{--}10^{-4}$), pinning the asymptotic spacetime suppression factor to $\mu_{\text{ST}} \to \mu_{T,\text{corr}}$ unless $\ket{3}$ leakage is actively mitigated.

Lastly, we confirm the model's prediction that the $\ket{3}$ state is limiting the logical error probability in the correlated error regime through direct measurement. 
Specifically, we run a short ($c = 14$ cycles) stability experiment on a $3\times3$ surface code patch and measure data qubits at the terminal layer of the circuit with a readout that resolves between the $\left\{\ket{0},\ket{1},\ket{2},\ket{3}\right\}$ states.
We investigate the final data-qubit state using data-qubit IQ values postselected on two conditions used to isolate temporal error bursts; specifically, we postselect on (i) logical failures and (ii) shots for which the detection fraction in the second half of the shot is higher than in the first half.
The latter condition allows us to nominally catch shots when the limiting correlated error chain runs into the end of the experiment.

The results of our direct data-qubit readout investigation are shown in \fig[d]{3}.
Notably, out of $N=21$ flagged events, a majority ($52\%$) terminate with the flagged data qubit in the third excited transmon state $\ket{3}$, whereas none terminate in $\ket{2}$, using simple readout classification.
This stark asymmetry between $\ket{2}$ and $\ket{3}$ arises from our use of $\ket{2}$-state DQLR in the experiment, without any engineered mechanism to remove the $\ket{3}$ state.
We also observe an outlier measurement, which may indicate the presence of even higher ($\ket{4+}$) data-qubit states and is the subject of ongoing investigation. 
Some leftover population in $\ket{0}$ and $\ket{1}$ is expected from our imperfect shot-picking heuristic; however, the strong presence of the data-qubit $\ket{3}$ state in shots with correlated error chains further confirms it as the limiting error mechanism in the correlated regime on this device, in addition to the agreement with the leakage model.

\section{Mitigating temporally correlated errors via walking} \label{sec:walking}

The \textit{walking} syndrome extraction circuit was introduced in \cite{mcewen_relaxing_2023} and experimentally realized for a memory experiment in \cite{Eickbusch2025}.
The walking circuit is a dynamic QEC protocol that alternates between two distinct cycle types.
This alternation enables a key property: by shuffling logical information between data and measure qubits, the circuit reads out and periodically resets all qubits.
As a result, when using a multi-level-reset scheme that resets higher qubit states (including $\ket{3}$), leakage is removed from all qubits periodically.
Given that the $\ket{3}$ state is limiting in the correlated error regime, a walking circuit can mitigate this floor by periodically removing the $\ket{3}$ state on all qubits when a multi-level reset is used.
In this work, we design a QEC circuit that realizes the walking stability code by modifying the boundaries of the memory walking circuit given in \cite{mcewen_relaxing_2023}.

\begin{figure}[t!]
    \centering
    \includegraphics[width=\columnwidth]{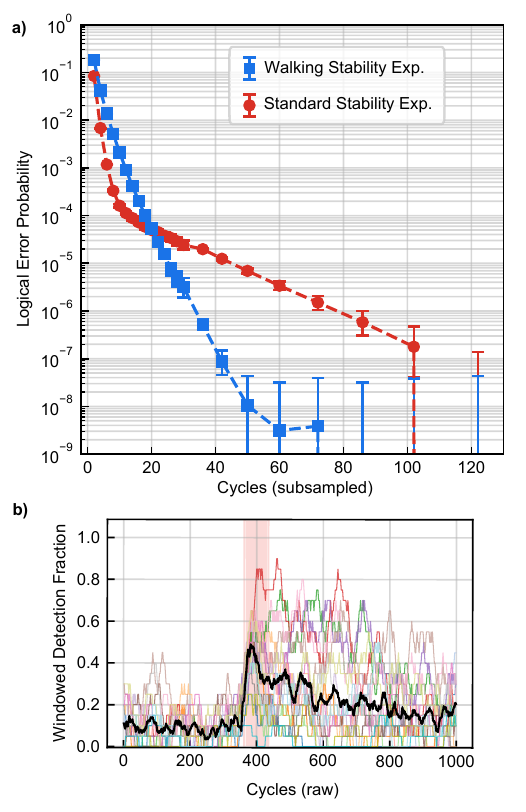}
    \caption{ \textbf{a)} Measured walking (blue) vs.\ standard non-walking (red) stability logical error on a $3\times3$ patch of surface code. 
    Walking stability data run a 999-cycle $3 \times 3$ $Z$-basis circuit on a single subgrid for $2 \times 10^{7}$ raw shots. Extracting non-overlapping subsampling windows yields $2 \times 10^{7}$ effective shots for $c \le 30$ (one window per shot) and $5.4 \times 10^{8}$ ($1.6 \times 10^{8}$) effective shots at $c = 36$ ($c = 122$) from $27$ ($8$) tiled windows per shot, with error bars spanning binomial probabilities within a factor of $1{,}000$ of the maximum likelihood (${\sim}3.7\sigma$, showing upper bounds where zero errors occur).
    \textbf{b)} Windowed detection fraction for the error burst causing the walking floor. The red region highlights a failed subsample.
    }
    \label{fig:4}
\end{figure}

We experimentally run the $3\times3$ walking stability circuit, with results compared against the standard (non-walking) stability circuit of the same size, as shown in \fig[a]{4}.
There is an expected overhead associated with walking given the larger qubit count, leading to a higher logical error probability in the Pauli regime. 
Here, the walking circuit lowers the suppression rate, as expected from a Pauli model simulation \suppsec{walking_stability}, yielding a logical error probability roughly an order of magnitude higher at 10 cycles. 
Although we incur this overhead, the walking circuit successfully mitigates the knee from the $\ket{3}$ state, reaching a logical error probability upper bound of $\le 3.20\times 10^{-8}$ at 60 cycles---a two-order-of-magnitude improvement over the non-walking circuit. 
Specifically, at $c = 60$ cycles, the walking stability circuit exhibits $1$ logical failure out of $3.2 \times 10^{8}$ non-overlapping subsampled shots ($3.1 \times 10^{-9}$ logical error probability), run over 5.55 hours of circuit execution time.

We find that, like the repetition code, the walking stability circuit is limited by ionizing radiation bursts \cite{Acharya2023, GoogleQuantumAI2024, mcewen_resolving_2022,McEwen2024resisting, Kurilovich2026}.
As shown in \fig[a]{4}, the walking stability floor is caused by a single error burst, with detection statistics for this burst shown in \fig[b]{4}.
It is likely that this second knee can be surpassed by extending the stability experiment beyond $\sim 1000$ cycles, since bursts typically last on a timescale of $\sim 1~\text{ms}$.

\section{Adaptive stopping with decoder confidence}

Unlike memory experiments, stability experiments allow protection against timelike errors to be chosen adaptively by deciding when to terminate \cite{akahoshi2026runtime}. In a fault-tolerant quantum computer, terminating early after clean cycles of syndrome extraction can accelerate logic, while delaying termination after noisy cycles can avoid timelike logical failures and substantially reduce the logical error rate. To make this decision quantitatively, we use a measure of decoder confidence called the complementary gap: the log-likelihood ratio of the two most likely sets of errors giving different observable outcomes as estimated by the decoder \cite{gidney_yoked_2023}. A gap of zero means the decoder is entirely unsure of its prediction---effectively a coin flip---while a large gap indicates high confidence.
Using an offline analysis, we compute the gap after each cycle, and terminate once it exceeds a syndrome-dependent threshold \suppsec{supp_complementary_gap}.

In a representative stability shot (\fig[a]{5}), we observe a sustained burst of detection events neighboring a single data qubit between cycles $6$ and $13$. This burst confuses the decoder, keeping the complementary gap small. Once the burst subsides, the decoder changes its prediction and, by cycle $20$, has become confident in the new one. With the resulting flagged threshold of $21$~nats, we wait out the burst and terminate with the correct prediction.
Adaptively deciding when to stop reduces the mean stopping time by $\approx 34\%$ and the spacetime logical error rate by $\approx 38\%$.
It has a considerably greater effect on the timelike logical error probability---a $>100\times$ reduction given a mean hold time of 12 cycles---but this offers diminishing returns when coupled with accumulating spacelike error at long hold times. Still, this dramatic reduction in timelike error suggests more significant spacetime logical error improvement as we move to larger distances in the presence of long temporally correlated error sources.

\begin{figure}[t!]
    \centering
    \includegraphics[width=\columnwidth]{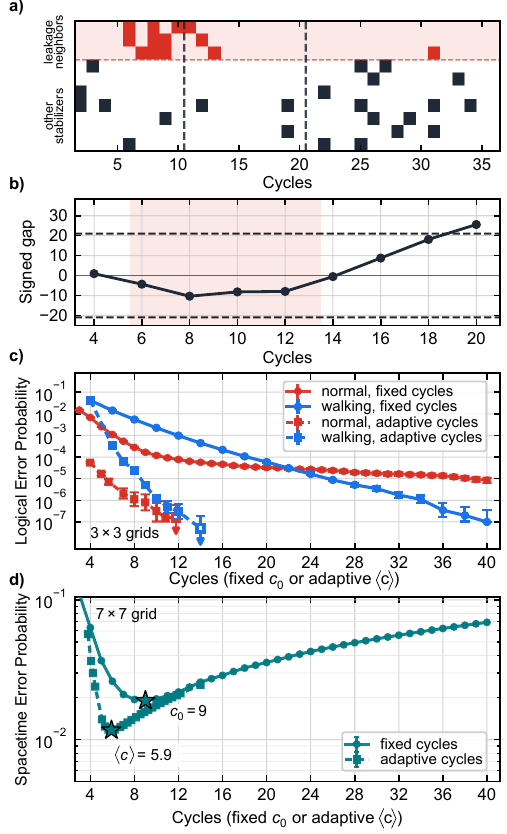}
    \caption{
    \textbf{a)} An example stability experiment. Activated detectors are highlighted. Measure qubits adjacent to a candidate leaked data qubit are highlighted in red.
    \textbf{b)} The random detection event pattern confuses the decoder, keeping its confidence low. After the pattern clears, the decoder gains confidence, eventually switching its decision (sign) and committing to it.
    \textbf{c)} Comparing fixed to adaptive strategies in the standard surface code and the walking surface code for a $3 \times 3$ patch.
    \textbf{d)} Estimating the spacetime logical error probability on the $7 \times 7$ grid by combining the timelike stability outcome of each shot at its stopping cycle $c$ with the accumulated spacelike memory error over $c$ cycles. With adaptive cycles, $50$\% of shots complete in $\leq 5$ cycles, $90$\% complete in $\leq 10$, and $99$\% in $\leq 18$.}
    \label{fig:5}
\end{figure}

The improvement in performance does not come for free. First, adaptively deciding when to stop round-to-round requires fast decoder decisions and feedback on the order of $1~\mu\text{s}$. This is conceivable but would be substantially alleviated by making decisions only every several cycles or using a simplified heuristic, both of which will affect performance. Also, while the mean hold time is reduced, there can be long hold times lying in the tail of the hold-time distribution.
Practically, compiling an algorithm with variable-length gates that can take quite long will be challenging.
Nevertheless, engineering the ability to wait out rare bad events may prove important for building a large-scale superconducting processor.

\section{Conclusion} \label{sec:conclusion}

In this work, we realized stability and memory experiments on a 105-qubit Willow processor to benchmark both spatial and temporal errors in quantum error correction. 
By combining memory and stability logical error probabilities into the spacetime error metric, we measure a spacetime suppression factor of $\mu_{\text{ST}} = 1.42 \pm 0.03$ as the patch size scales from $3\times 3$ to $7\times 7$ in the low-cycle Pauli regime. 
Moving forward, device performance improvements will improve this suppression factor, and we estimate that using IQ data (soft information) in decoding, along with a more sophisticated decoder, will improve the stability performance \cite{Ali2024, Bausch2024learning}. 

Extending the stability experiment to higher cycle counts reveals a sharp knee in logical error probability caused by data-qubit heating to the third excited state $\ket{3}$. 
Because our data-qubit leakage removal (DQLR) operation is calibrated to remove $\ket{2}$-state leakage via a leakage-iSWAP interaction without targeting the $\ket{3}$ state, data qubits that heat to $\ket{3}$ remain in $\ket{3}$ for roughly ten cycles until cleared by intrinsic relaxation or leakage transport to measure qubits during CZ gates, reducing the measured Pauli-regime temporal suppression factor from the Pauli-model prediction of $\mu_T = 1.97$ to $1.47$. 
A device-level multi-level simulation with no free parameters reproduces this knee across all patch widths and shows that, if left unmitigated, $\ket{3}$-state leakage will overtake spatial memory errors at code distances $d^* \approx 11\text{--}13$. 
Finally, by implementing a walking stability circuit in which every qubit alternates between data and measure roles and undergoes a multi-level reset (which resets higher excited states including $\ket{3}$) every two cycles, we remove the $\ket{3}$-state knee and reach a stability logical error probability floor of $\le 3.20 \times 10^{-8}$ at 60 cycles, limited by ionizing radiation error bursts and suppressing logical error by two orders of magnitude compared to the standard compilation at 60 cycles.

While the walking circuit eliminates the $\ket{3}$-state knee, it introduces higher logical error rates at low cycle counts than the standard circuit (\fig[a]{4}). 
Several alternative strategies could mitigate $\ket{3}$-state leakage without walking the surface code patch. 
First, DQLR could be extended to also transfer $\ket{3}$-state population from data qubits to measure qubits prior to measure-qubit reset, while leaving the computational subspace undisturbed. 
Second, reducing passive heating rates into $\ket{3}$ during data-qubit idling and dynamical decoupling would directly lower the onset probability of the knee. 
Third, at the decoding level, monitoring the complementary gap after each cycle allows the duration of a stability experiment to be chosen adaptively---terminating early after clean cycles or running additional cycles to wait out temporal error chains---yielding considerable improvement in timelike logical error (\fig{5}).

Together, these results show that memory experiments alone are not sufficient to predict the performance of fault-tolerant logical operations. 
Primitive operations such as lattice surgery and logical qubit moves require $c \approx d$ cycles of stabilizer measurement and contain spacetime regions sensitive to both spatial and temporal errors. 
In the Pauli regime, our spacetime error measurements indicate that logical operations benefit from waiting $c^* \approx d+1$ to $d+2$ cycles rather than $c = d$, while our error budgets show that mid-circuit readout ($4.28\times$) and reset ($2.43\times$) carry substantially higher relative weight in stability compared to memory (\fig[e,f]{2}). 
More broadly, because the fault distance of a stability experiment is set by the number of cycles ($d = c$) rather than the number of physical qubits, even small surface code patches can probe temporal error floors down to $10^{-8}$, providing a practical metrology tool for identifying and eliminating correlated error mechanisms before scaling to larger fault-tolerant processors.

\textit{Note added.}---While finalizing this manuscript, we became aware of \cite{Harper2026entangled}, which realizes a similar circuit construction to \sec{walking}.

\vspace{-3mm}
\section*{Contributions}
\vspace{-1mm}

A. Eickbusch led the project, performed the experiments, analyzed the data, developed the simple modeling, and realized the walking code. N. Noll led the modeling effort, developed the detailed modeling, built the leakage simulation, and analyzed the impact of leakage to higher excited states as a time-correlated error. A. Eickbusch and N. Noll conceived of the project. A. Eickbusch, N. Noll, and M. Newman wrote the paper with input from all the authors, and A. Morvan provided support on the manuscript. M. Newman created the gap figures and adaptive stopping analysis, wrote the corresponding supplement, and together with C. Gidney provided overall guidance throughout the project. M. McEwen guided the spacetime logical error analysis. L. Martin developed and implemented echoes in the walking code. G. Roberts, N. Lacroix, A. Bourassa, E. Rosenfeld, A. Bengtsson, K. Satzinger, and R. Cortiñas assisted in code infrastructure development and calibrations. V. Kurilovich, Y.-C. Wei, and G. Roberts assisted with the analysis of error bursts. C. Ding, L. De Lorenzo, and F. Borjans developed infrastructure and measured heating rates and population transport. V. Sivak developed the reinforcement learning calibration. A. Korotkov, S. Boixo, K. Miao, A. Opremcak, Z. Chen, C. Jones, J. Kelly, D. Kafri, and C. Neill managed the teams and development, and Y. Chen oversaw the project. H. Neven created an environment where this research was possible. All authors contributed to building the hardware and software, and writing the paper.

\vspace{-3mm}
\section*{Acknowledgements}
\vspace{-1mm}
We thank Daniel Sank for valuable discussions and feedback on this manuscript. 
\vspace{-3mm}
\section*{Corresponding authors}
\vspace{-1mm}
Correspondence and requests for materials should be addressed to Alec Eickbusch (aleceickbusch@google.com) and Nick Noll (nicknoll@google.com).

\vspace{-3mm}
\section*{Competing Interests}
\vspace{-1mm}
The authors declare no competing interests.

\vspace{-3mm}
\section*{Code and data availability}
\vspace{-1mm}
The datasets generated and analyzed for this study are available at 
\url{https://doi.org/10.5281/zenodo.24142014}

\newpage

\onecolumngrid

\vspace{1em}

\begin{flushleft}

{\small \textbf{Google Quantum AI and Collaborators}}

\bigskip

{\small

\renewcommand{\author}[2]{#1$^\textrm{\scriptsize #2}$}

\renewcommand{\affiliation}[2]{$^\textrm{\scriptsize #1}$ #2 \\}

\newcommand{\corrauthora}[2]{#1$^{\textrm{\scriptsize #2}, \ddagger}$}

\newcommand{\xGoogle}{\affiliation{1}{Google Quantum AI, Santa Barbara, CA 93117, USA}}
\newcommand{\xUMass}{\affiliation{2}{Department of Electrical and Computer Engineering, University of Massachusetts, Amherst, MA, USA}}
\newcommand{\xUCSB}{\affiliation{3}{Department of Physics, University of California, Santa Barbara, CA, USA}}
\newcommand{\xAuburn}{\affiliation{4}{Department of Electrical and Computer Engineering, Auburn University, Auburn, AL, USA}}
\newcommand{\xUConn}{\affiliation{5}{Department of Physics, University of Connecticut, Storrs, CT, USA}}

\newcommand{\Google}{1}
\newcommand{\UMass}{2}
\newcommand{\UCSB}{3}
\newcommand{\Auburn}{4}
\newcommand{\UConn}{5}

\corrauthora{Alec Eickbusch}{\Google},
\corrauthora{Nick Noll}{\Google},
\author{Michael Newman}{\Google},
\author{Matt McEwen}{\Google},
\author{Leigh S.~Martin}{\Google},
\author{Alexis Morvan}{\Google},
\author{Gabrielle Roberts}{\Google},
\author{Chunyang Ding}{\Google},
\author{Volodymyr Sivak}{\Google},
\author{Nathan Lacroix}{\Google},
\author{Laura De~Lorenzo}{\Google},
\author{Felix Borjans}{\Google},
% ...

\bigskip

\xGoogle
\xUMass
\xUCSB
\xAuburn
\xUConn

{${}^\ddagger$ These authors contributed equally to this work.}

}

\end{flushleft}

\bibliography{bib}

\FloatBarrier
\clearpage
\makeatletter
\expandafter\def\expandafter\section\expandafter{\expandafter\FloatBarrier\section}
\makeatother
\onecolumngrid

\begin{center}
    {\large \bfseries Supplementary Material for: \\[0.5em]
    \PaperTitle} \\[1.5em]
    
    {\normalsize \PaperAuthors}
\end{center}

\vspace{2em}

\setcounter{equation}{0}
\setcounter{figure}{0}
\setcounter{table}{0}
\setcounter{page}{1}
\setcounter{section}{0}

\renewcommand{\theequation}{S\arabic{equation}}
\renewcommand{\thefigure}{S\arabic{figure}}
\renewcommand{\thetable}{S\arabic{table}}
\renewcommand{\thepage}{S\arabic{page}}
\renewcommand{\thesection}{S\arabic{section}}
\renewcommand{\theHequation}{S\arabic{equation}}
\renewcommand{\theHfigure}{S\arabic{figure}}
\renewcommand{\theHtable}{S\arabic{table}}
\renewcommand{\theHsection}{S\arabic{section}}

\section{Stability experiment layout and definitions} 
\label{sec:stability_supplement}

A stability experiment encodes zero logical qubits and tracks a global stabilizer product across time to isolate timelike measurement and ancilla errors.
Surrounding a patch entirely with boundaries of one type ($X$ or $Z$) cancels all data-qubit operators in the product of stabilizers of that type ($S$), defining a global parity observable $L_0 = \prod_{i}S_i = 1$ that is constant across error-free cycles.
This can be seen by noticing that all data qubits enter into the product an even number of times.

\begin{figure}[!htbp]
  \centering
  \includegraphics[width=0.8\textwidth]{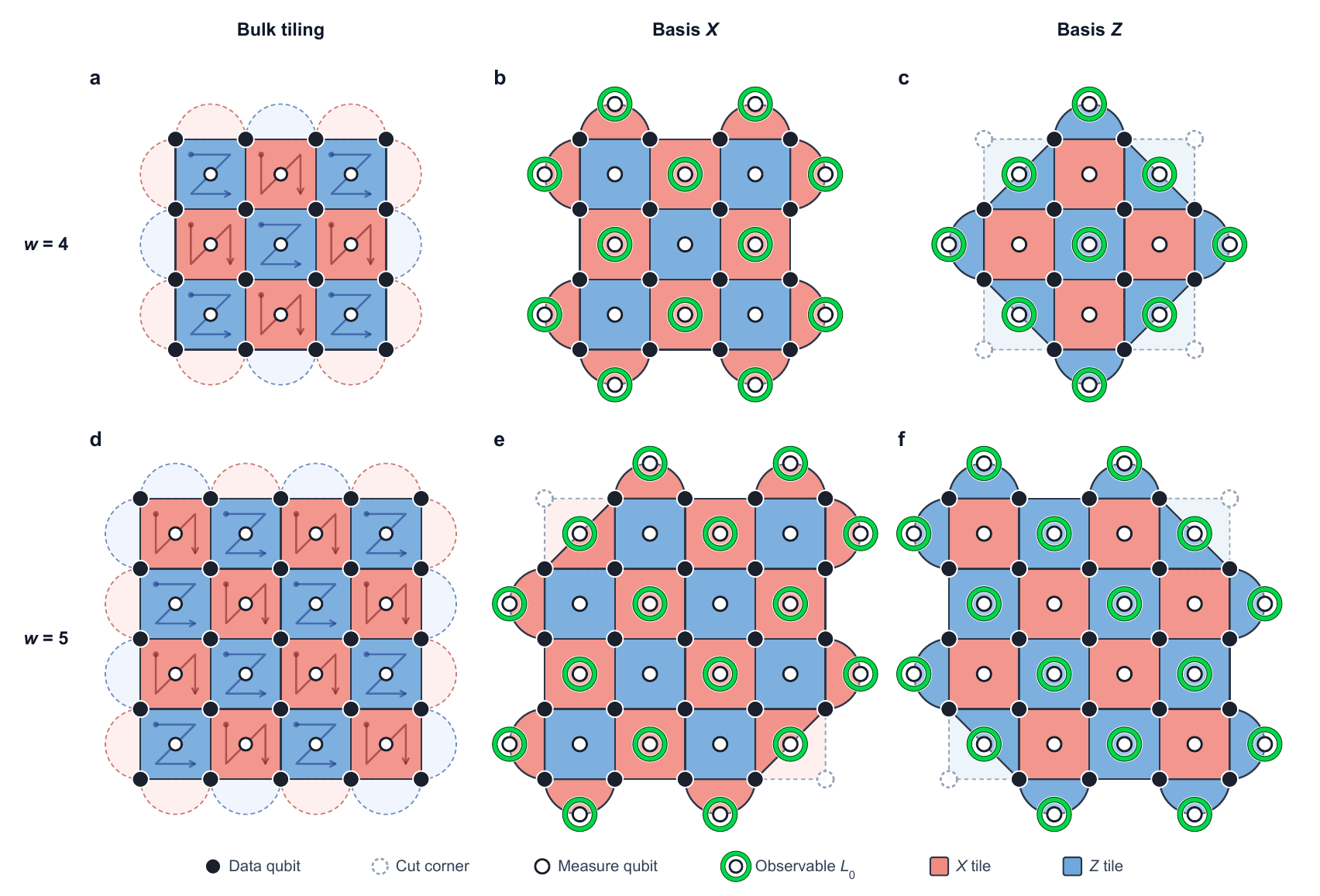}
  \caption{\textbf{Stability patch construction across width and basis.}
  Patches for $w = 4$ (\textbf{a--c}) and $w = 5$ (\textbf{d--f}) on a fixed bulk checkerboard.
  \textbf{a, d,} Bulk tiling with interior arrows showing the two-qubit $\mathrm{CZ}$ gate order and dashed arcs showing candidate weight-2 boundaries.
  \textbf{b, e,} Basis $X$ and \textbf{c, f,} Basis $Z$ patches; green circles mark measure qubits included in $L_0$, and dashed circles mark removed corner data qubits.}
  \label{fig:patch_parameterization}
\end{figure}

Each stability patch is specified by its width $w$, a fixed bulk checkerboard tiling, and the boundary basis ($X$ or $Z$; Fig.~\ref{fig:patch_parameterization}). 
Defining the bulk tiling first locks the four-layer $\mathrm{CZ}$ gate order across $X$-stability, $Z$-stability, and paired memory runs, after which choosing the basis keeps only the matching weight-2 perimeter checks. Wherever a corner bulk tile matches the boundary basis, the adjacent perimeter checks are absent, so that corner data qubit belongs to only one check in $L_0$ and is removed to leave a weight-3 corner check. Even widths ($w = 4$) therefore keep all four corners in one basis and cut all four in the other, whereas odd widths ($w = 5$) cut opposite pairs of corners.

\begin{figure}[!htbp]
  \centering
  \includegraphics[width=0.8\textwidth]{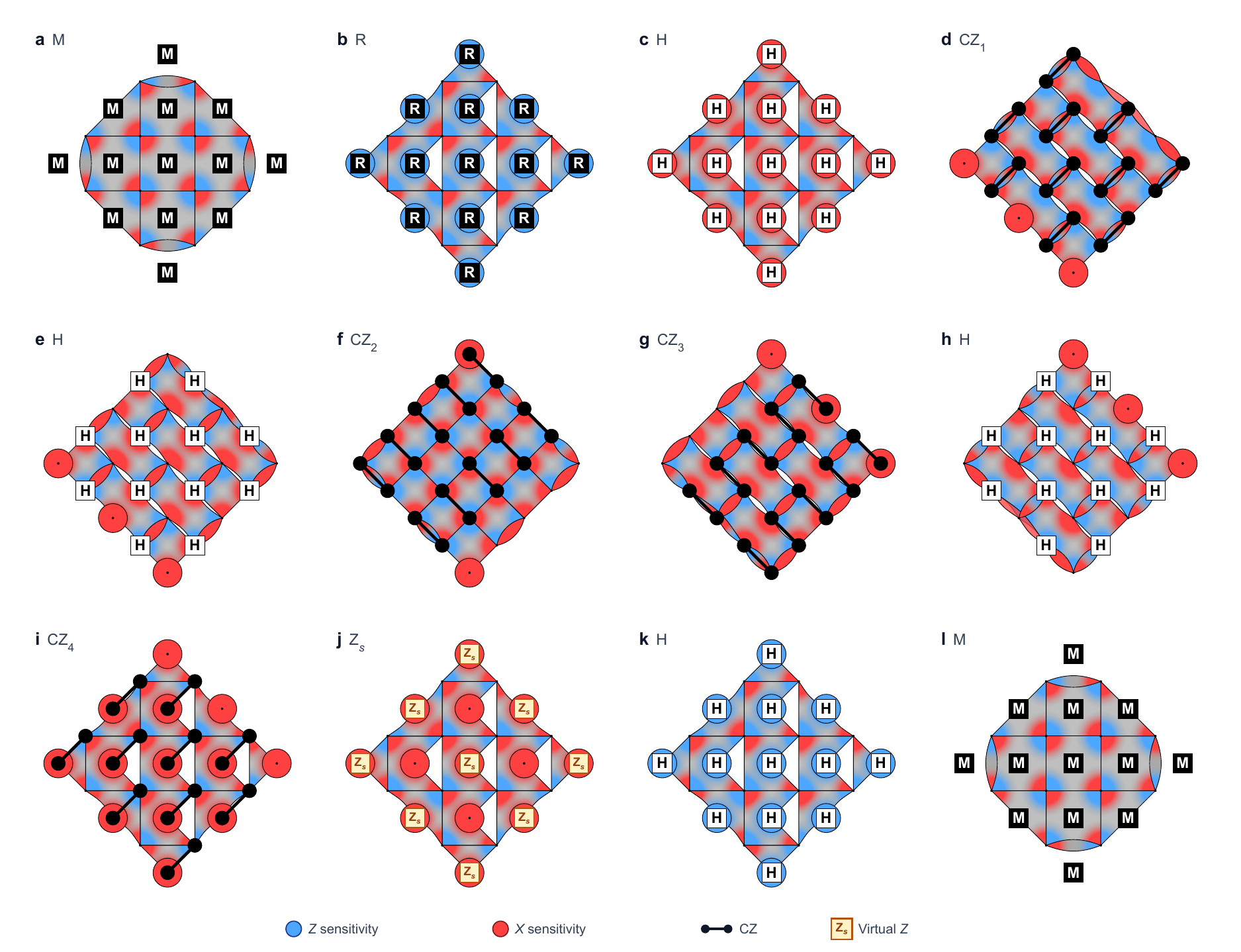}
  \caption{\textbf{Single-cycle $\mathrm{M}$-to-$\mathrm{M}$ detector slice of the ZXXZ stability circuit.}
  Steady-state cycle for the $w = 4$, Basis $Z$ patch of Fig.~\ref{fig:patch_parameterization}c (blue: $Z$ sensitivity; red: $X$ sensitivity). Ancilla measurements $\mathrm{M}$ (\textbf{a, l}) close and reopen the ZXXZ detector slice; reset and Hadamard (\textbf{b, c}) precede four $\mathrm{CZ}$ layers (\textbf{d, f, g, i}) interleaved with data Hadamards (\textbf{e, h}); and classical-bit-controlled virtual $Z$ gates ($Z_s$, \textbf{j}) precede the readout Hadamards (\textbf{k}).}
  \label{fig:detslice}
\end{figure}

Data qubits are initialized in $\ket{+Y}$ and measured in the $Y$ basis in the first and last cycles.
Since single-qubit $Y$ states anticommute with individual $X$ and $Z$ checks, all individual initial and final stabilizer outcomes are randomized while their global product $L_0$ remains deterministic.
Before the measure-qubit readout Hadamard in each cycle, virtual $Z$ gates ($Z_s$ in Fig.~\ref{fig:detslice}j) are randomly applied, realizing random per-shot sign flips to the ancillas in $L_0$.
Keeping these classical bits fixed across cycles leaves two-cycle detectors unchanged while randomizing $L_0$ from shot to shot.

\begin{figure}[!htbp]
  \centering
  \includegraphics[width=0.8\textwidth]{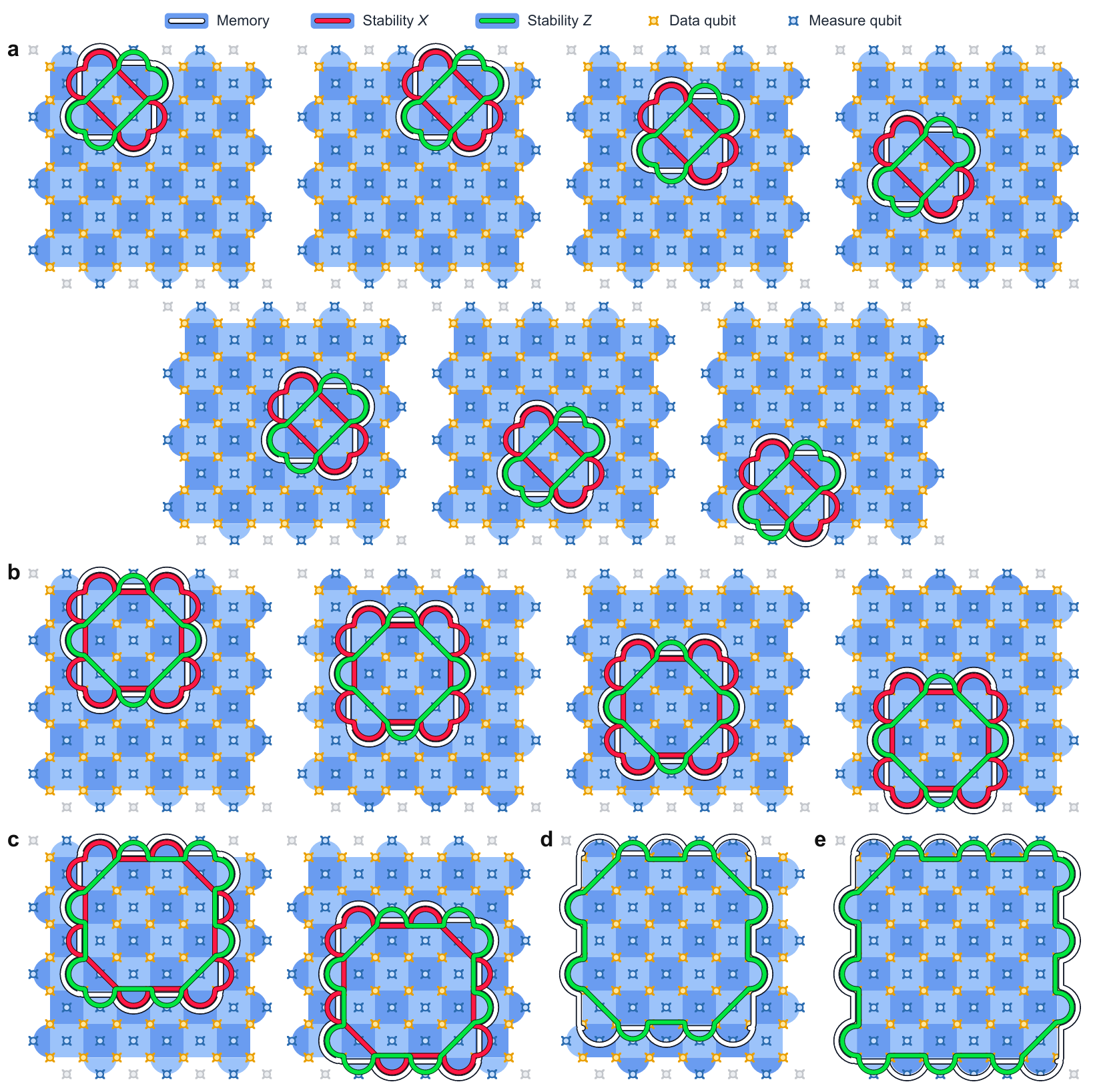}
  \caption{\textbf{Memory and stability subpatch tilings on a 105-qubit superconducting processor.}
  Outlines show Memory ($X$ and $Z$; white), Stability $X$ (red), and Stability $Z$ (green) across \textbf{a,} seven $d = 3$, \textbf{b,} four $d = 4$, \textbf{c,} two $d = 5$, \textbf{d,} one $d = 6$, and \textbf{e,} one full-chip $d = 7$ subpatch (gray crosses mark unused peripheral qubits; darker and lighter blue tiles show local $X$ and $Z$ plaquette sublattices).}
  \label{fig:processor_tiling}
\end{figure}

We run paired memory and stability experiments across 15 subpatches spanning $d \in \{3, 4, 5, 6, 7\}$ on a 105-qubit superconducting processor (Fig.~\ref{fig:processor_tiling}). At $d = 3, 4,$ and $5$ (Fig.~\ref{fig:processor_tiling}a--c), subpatches sit away from the left and right chip edges so that all perimeter ancillas exist, allowing Memory ($X$ and $Z$; white), Stability $X$ (red), and Stability $Z$ (green) to run on every location. At $d = 6$ and $d = 7$ (Fig.~\ref{fig:processor_tiling}d,e), peripheral ancillas on the top and bottom $Z$ boundaries support Stability $Z$ alongside Memory $X$ and $Z$, whereas the left and right edges carry no extra $X$-boundary ancillas, so Stability $X$ is omitted. On hardware, a few additional neighboring qubits outside each patch are also used for data-qubit leakage removal (DQLR) on boundary data qubits.

\section{Heuristic model including temporally correlated errors}
\label{sec:correlated_stability_model}

In a stability experiment of $c$ cycles, a timelike logical error occurs when errors accumulate on at least $c/2$ cycles along a timelike path, causing a minimum-weight decoder to match to the $t = 0$ and $t = c$ time boundaries instead of pairing errors locally. With uncorrelated Pauli errors alone, the logical error probability decays exponentially with $c$ as $P_{L, 1}(c) = a_1 s_1^{-c}$, where $s_1 > 1$ is the per-cycle Pauli suppression factor and $a_1$ captures boundary effects and spatial path counting across the patch.

To model the knee at larger $c$, suppose a rare correlated error occurs anywhere in the patch with per-shot probability $p_0 \ll 1$ (and does not occur with probability $1 - p_0$), producing a contiguous timelike error chain of length $N \in \{1, 2, \dots\}$ cycles. For physical mechanisms with memory that have a constant probability per cycle of relaxing or being reset, such as leakage to higher transmon states or coupling to a two-level system (TLS) with a finite lifetime $T_1$, the chain length follows an exponential distribution,
\begin{equation}
    P(N) = p_0 \left(e^{1/\tau} - 1\right) e^{-N/\tau},
    \label{eq:exp_chain_pmf}
\end{equation}
where $\tau$ is the characteristic correlation time in cycles and $\sum_{N=1}^{\infty} P(N) = p_0 \ll 1$ (since $P(N)$ counts only the correlated error chains, while $1 - p_0$ is the probability of having none).

If a chain triggers detection events only at its start and end cycles, any single chain with $N \ge c/2$ spans half the experiment and causes a logical error on its own. Neglecting combinations of shorter chains with Pauli errors ($s_1 \gg s_2$) and taking $c$ even for simplicity, summing $P(N)$ over $N \ge c/2$ gives $P_{L, \mathrm{corr}}(c) \approx p_0 e^{1/\tau} e^{-c/(2\tau)}$. Adding this to the Pauli term $(1 - p_0) P_{L, 1}(c)$ gives the heuristic sum-of-exponentials model used in the main text,
\begin{equation}
    P_L(c) = a_1 s_1^{-c} + a_2 s_2^{-c},
    \label{eq:double_exp_stability}
\end{equation}
with
\begin{align}
    s_2 &= e^{1/(2\tau)}, \label{eq:s2_relation} \\
    a_2 &= p_0 e^{1/\tau} = p_0 s_2^2, \label{eq:a2_relation}
\end{align}
so $a_2 \approx p_0$ for $\tau \gg 1$. Whenever correlated chains decay more slowly than Pauli errors ($s_2 < s_1$), $P_L(c)$ bends at a knee from steep Pauli suppression to a shallower tail set by $\tau$.

In practice, a leaked qubit or active TLS can also trigger detection events during the middle of the chain with some probability $\zeta \in [0, 1)$ per cycle, allowing the decoder to pair segments locally in the bulk rather than to the time boundaries. 
In addition, since this simple picture also leaves out starting-time multiplicity, hybrid Pauli-correlated chains, and three-dimensional decoding details, we treat Eq.~\eqref{eq:double_exp_stability} as a phenomenological fit and use circuit-level simulations to study specific error mechanisms.

\subsection{Chain-length distributions, staircases, and multiple knees} \label{sec:supp_non_markovian_models}

\begin{figure}[!htbp]
    \centering
    \includegraphics[width=\textwidth]{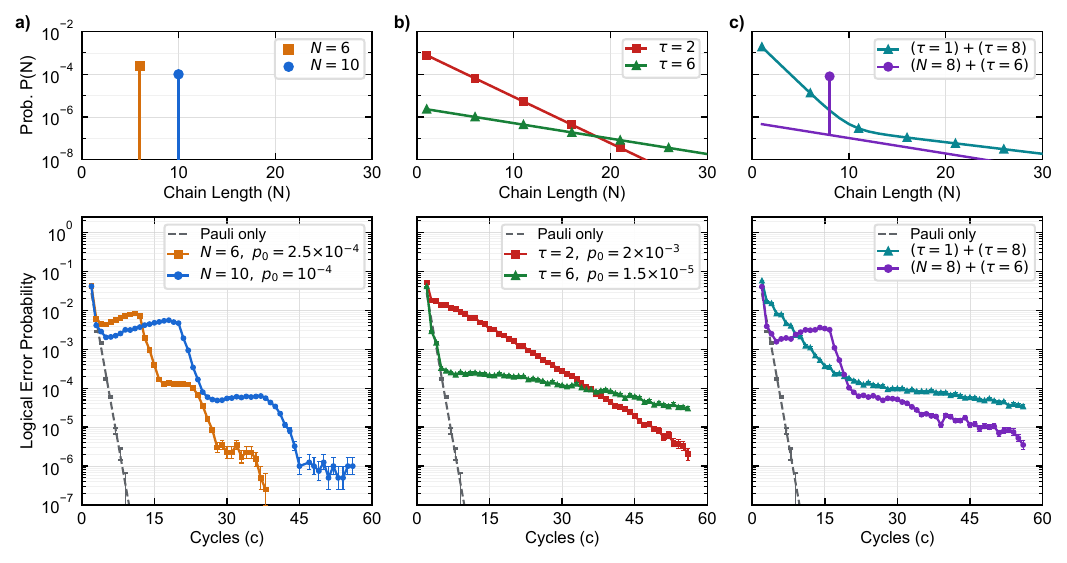}
    \caption{\textbf{Simulated stability experiments under different correlated error models.}
    Top row: correlated chain-length probability $P(N)$ per stabilizer and per cycle versus chain length $N$, with $\sum_{N=1}^{\infty} P(N) = p_0 \ll 1$. Bottom row: simulated stability logical error probability $P_L(c)$ versus cycles $c$ for a $3 \times 3$ $Z$-basis ZXXZ surface code patch under $\mathrm{SI1000}$ noise ($p = 10^{-3}$), decoded with correlated minimum-weight perfect matching. Correlated errors are added after circuit sampling by flipping a stabilizer's measurement outcomes for $N$ consecutive cycles.
    \textbf{a)} Fixed-length chains, $P(N) = p_0 \delta_{N, N_0}$, for $N = 6$ ($p_0 = 2.5 \times 10^{-4}$, orange squares) and $N = 10$ ($p_0 = 10^{-4}$, blue circles). Within each step, $P_L(c)$ rises with $c$ because a longer experiment provides more opportunities in time for a limiting error chain to occur, before dropping sharply at multiples of $2N$ once an additional chain is needed to span half the experiment.
    \textbf{b)} Exponentially distributed chains, $P(N) = p_0 (e^{1/\tau} - 1)e^{-N/\tau}$, consistent with leakage or coupling to a TLS with finite lifetime $T_1$, shown for $(\tau = 2, p_0 = 2 \times 10^{-3})$ (red squares) and $(\tau = 6, p_0 = 1.5 \times 10^{-5})$ (green triangles). Both decay smoothly after the knee and cross near $c \approx 36$, where the longer-lived error takes over.
    \textbf{c)} Combined models with multiple knees: combining two exponentials with $(\tau_1 = 1, p_1 = 3 \times 10^{-3})$ and $(\tau_2 = 8, p_2 = 6 \times 10^{-6})$ (cyan triangles) produces two knees separating three monotonic decay slopes, while combining fixed-length chains ($N = 8$, $p_{\mathrm{fixed}} = 8 \times 10^{-5}$) with an exponential tail ($\tau = 6$, $p_{\mathrm{exp}} = 3 \times 10^{-6}$) (purple circles) produces a rising plateau up to $c = 16$ followed by a drop into the $\tau = 6$ tail.}
    \label{fig:supp_non_markovian_models}
\end{figure}

An exponential chain-length distribution is most consistent with leakage and other physical temporally correlated error mechanisms with memory, such as coupling to a TLS with a finite lifetime $T_1$, and produces a single smooth decay after the knee. Different correlated error processes, however, can have different chain-length distributions $P(N)$, producing staircases, regions where the logical error worsens with cycles, or multiple knees (Fig.~\ref{fig:supp_non_markovian_models}).

To build intuition for these features, we simulate a $3 \times 3$ stability experiment and inject correlated errors by flipping a stabilizer's measurement outcomes for $N$ consecutive cycles with probability $p_0 = \sum_{N=1}^{\infty} P(N) \ll 1$ per cycle. When chains have a fixed length $N$ (Fig.~\ref{fig:supp_non_markovian_models}\textbf{a}), a single chain is long enough to cause a logical error whenever $c \le 2N$. Within this window, running the experiment for more cycles gives more opportunities in time for such a limiting error to occur, a temporal path-counting effect analogous to the spatial path counting in the main text, causing $P_L(c)$ to rise with $c$. Once $c > 2N$, a single chain is too short to span half the experiment on its own and two chains are required, causing a sharp drop onto the next step and forming a staircase with period $2N$. For an exponential distribution (Fig.~\ref{fig:supp_non_markovian_models}\textbf{b}), longer chains are exponentially rarer, overcoming path counting to give a smooth post-knee decay where longer-lived errors eventually overtake shorter-lived ones at large $c$. Combining mechanisms (Fig.~\ref{fig:supp_non_markovian_models}\textbf{c}) produces multiple knees, either bending smoothly between distinct exponential slopes or rising along a step before dropping into an exponential tail.

\section{Numerical model} \label{sec:supp_numerical_model}

To capture both stochastic Pauli errors and multi-cycle temporal correlations arising from higher excited transmon states, we simulate our experiments using two complementary frameworks: (i) a circuit-level stochastic Pauli model (implemented in \texttt{Stim}~\cite{gidney_stim_2021}) for rapid detector statistics and Pauli-regime error budgeting, and (ii) a multi-level stabilizer-frame simulator tracking states $\ket{0}, \ket{1}, \ket{2},$ and $\ket{3}$ for full leakage dynamics.

During data-qubit idling and dynamical decoupling (DD) windows (which coincide with measure-qubit readout and reset), we model incoherent population transfer among the lowest four transmon levels $\ket{0}\text{--}\ket{3}$ as a continuous-time Markov process parameterized by empirical transition times $T_{i\to j} = \Gamma_{i\to j}^{-1}$. 
Fundamental relaxation ($T_{1\to 0}$) and dephasing ($T_{\phi}$) rates are assigned per qubit from individual idle and DD benchmarks, while higher-level relaxation times ($T_{2\to 1} = 36.57~\mu\text{s}$, $T_{3\to 2} = 21.81~\mu\text{s}$), single-photon heating times ($T_{0\to 1} = 4.91~\text{ms}$, $T_{1\to 2} = 2.60~\text{ms}$, $T_{2\to 3} = 1.47~\text{ms}$), and direct two-photon heating times ($T_{0\to 2} = 135.36~\text{ms}$, $T_{1\to 3} = 34.64~\text{ms}$) are applied uniformly across the grid based on device characterization.
At the end of each syndrome extraction cycle, Data Qubit Leakage Removal (DQLR)~\cite{Miao2023} is applied to all data qubits. 

We model DQLR as a state-selective stochastic reset channel on leaked data qubits that returns $\ket{2} \to \ket{1}$ with probability $\eta_{\ket{2}} = 0.98$, while clearing $\ket{3}$ with probability $\eta_{\ket{3}} = 0.06$, set to the median characterized values from the device. 
Measure qubits undergo a multi-level reset each cycle that clears both $\ket{2}$ and $\ket{3}$ back to the computational subspace, up to the benchmarked reset error $p_{\mathrm{reset}}$. 
We assume all leaked states are read out as ``1'' in the simulation.

The residence time of a data qubit in $\ket{3}$ is governed by the interplay between intrinsic relaxation ($T_{3\to 2} = 21.81~\mu\text{s}$) and population transfer during two-qubit CZ gates (``leakage transport''). 
In order to model multi-level leakage transport during CZ gates accurately without full waveform-level master-equation integration across the grid, we construct an effective block-diagonal unitary channel for each interacting qubit pair from a coupled transmon--coupler--transmon Duffing oscillator model parameterized by the calibrated idle frequencies and measured anharmonicities of both qubits, along with the calibrated interaction frequency and hold time of the gate. 
Leakage transport is mediated when a data qubit in $\ket{3}$ participates in a CZ gate with a measure qubit in $\ket{0}$ ($N=3$ manifold spanned by $\{\ket{30}, \ket{21}, \ket{12}, \ket{03}\}$) or $\ket{1}$ ($N=4$ manifold spanned by $\{\ket{40}, \ket{31}, \ket{22}, \ket{13}, \ket{04}\}$): Bose-enhanced off-resonant exchange transfers excitation from the data qubit onto the measure qubit (e.g., $\ket{30} \to \ket{21}, \ket{12}$ or $\ket{31} \to \ket{22}$). 
As each bulk data qubit executes four CZ gates per cycle with neighboring measure qubits (which are reset every cycle) and undergoes DQLR (which clears any resulting $\ket{2}$ population on the data qubit with $98\%$ probability), this CZ-mediated transport provides the primary removal pathway for data-qubit $\ket{3}$ excitations, setting the post-knee stability decay rate observed in \fig[a]{3}.

\section{Component error rates} 
\label{sec:supp_component_error_rates}

\begin{figure*}[t]
    \centering
    \includegraphics[width=\textwidth]{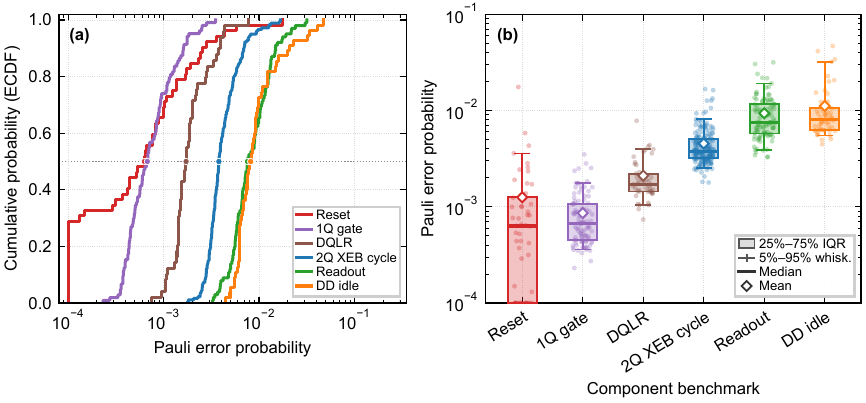}
    \caption{\textbf{Distributions of directly measured component Pauli error probabilities entering the surface code simulation.}
    \textbf{(a)}~Empirical cumulative distribution functions (ECDFs) of the per-target Pauli error probabilities across the processor; circular markers along the dotted line ($\mathrm{ECDF}=0.5$) indicate the median of each distribution.
    The noise floor of the reset benchmark caps the distribution at Pauli error $<1\times 10^{-4}$.
    \textbf{(b)}~Corresponding boxplots for each component benchmark on a logarithmic vertical scale, showing the interquartile range (25th--75th percentiles, shaded boxes), median (horizontal solid line), 5th--95th percentile range (whiskers), mean (open diamonds), and individual target measurements (points).}
    \label{fig:supp_component_error_rates}
\end{figure*}

To calibrate the circuit-level Pauli noise model used in our surface code simulations, we benchmark single-qubit (1Q) and two-qubit (2Q) operations in parallel circuit layers matching the syndrome-extraction schedule. Figure~\ref{fig:supp_component_error_rates} shows the empirical cumulative distribution functions (ECDFs) and boxplots of the directly measured component error rates across the Willow processor, and Table~\ref{tab:component_error_rates} summarizes the mean, median, and standard deviation of each distribution. All error rates are reported in units of Pauli error probability $p$, defined as the total probability of any non-identity Pauli error (related to the average gate infidelity $e = 1 - F_{\mathrm{avg}}$ in Hilbert-space dimension $d = 2^n$ via $p = \frac{d + 1}{d} e$).

Specifically, simultaneous 1Q Clifford randomized benchmarking (RB) measures the 1Q gate infidelity $e_{1\mathrm{Q}}$ ($p_{1\mathrm{Q}} = \frac{3}{2} e_{1\mathrm{Q}}$), while parallel 2Q cross-entropy benchmarking (XEB) measures the cycle infidelity $e_{\mathrm{cycle}}$ of two parallel 1Q gates followed by a controlled-$Z$ (CZ) gate ($p_{\mathrm{XEB}} = \frac{5}{4} e_{\mathrm{cycle}}$; in circuit simulations where 1Q gate errors are applied separately, the isolated CZ gate error is obtained by subtracting the two constituent 1Q gate errors, $p_{\mathrm{CZ}} \approx p_{\mathrm{XEB}} - p_{1\mathrm{Q},a} - p_{1\mathrm{Q},b}$). Interleaved 1Q RB directly measures the Pauli error $p_{\mathrm{DQLR}}$ of the data-qubit leakage removal (DQLR) operation, and state-preparation-and-measurement benchmarks characterize the residual excited-state probability after reset ($p_{\mathrm{reset}} = e_{\mathrm{reset}}$) and the state-averaged readout assignment error ($p_{\mathrm{RO}} = \frac{1}{2}[P(1|0) + P(0|1)]$), both modeled as bit-flip ($X$) errors. Finally, while ancilla qubits undergo measurement and reset, data qubits idle under dynamical decoupling (DD), where longitudinal and transverse purity decay benchmarks measure the per-cycle relaxation and dephasing probabilities $e_{T_1}$ and $e_{T_2}$ (giving a total Pauli error $p_{\mathrm{DD}} = \frac{e_{T_1}}{2} + \max(0, \frac{e_{T_2}}{2} - \frac{e_{T_1}}{4})$ under Pauli twirling).

\begin{table}[t]
\centering
\small
\setlength{\tabcolsep}{6pt}
\begin{tabular}{l c r r r}
\toprule
Component & $N$ & Mean & Median & Std.\ Dev. \\
\midrule
Reset error ($p_{\mathrm{reset}}$) & 52 & $0.125\%$ & $0.063\%$ & $0.257\%$ \\
1Q gate error ($p_{1\mathrm{Q}}$) & 104 & $0.086\%$ & $0.067\%$ & $0.058\%$ \\
Data-qubit leakage removal (DQLR) error ($p_{\mathrm{DQLR}}$) & 49 & $0.211\%$ & $0.172\%$ & $0.122\%$ \\
2Q XEB cycle error ($p_{\mathrm{XEB}}$) & 180 & $0.454\%$ & $0.380\%$ & $0.230\%$ \\
Readout error ($p_{\mathrm{RO}}$) & 101 & $0.938\%$ & $0.757\%$ & $0.524\%$ \\
Dynamical decoupling (DD) idle error ($p_{\mathrm{DD}}$) & 55 & $1.115\%$ & $0.816\%$ & $0.874\%$ \\
\bottomrule
\end{tabular}
\caption{Mean, median, and standard deviation of the directly measured component Pauli error probabilities ($\%$) across $N$ benchmarked targets entering the surface code simulation noise model.}
\label{tab:component_error_rates}
\end{table}

\section{Error budgeting} \label{sec:supp_error_budgeting}

To construct the Pauli-regime error budgets shown in \fig[e,f]{2}, we simulate memory, stability, and combined spacetime logical error probabilities across odd code distances $d \in \{3, 5, 7\}$ using our benchmarked circuit-level Pauli noise model (Sec.~\ref{sec:supp_component_error_rates}) and ablate each error channel family one at a time. The optimal memory error $P^*_M = P_M(c^*)$, stability error $P^*_S = P_S(c^*)$, and minimum spacetime error $P^*_{\mathrm{ST}}(c^*) = \frac{1}{2}(P^*_M + P^*_S)$ are evaluated on a $w \times w \times c^*$ spacetime block ($w = d$) at the fixed experimental optimal cycle counts $c^* \in \{4, 7, 9\}$ for $d \in \{3, 5, 7\}$ (held constant across the baseline and all single-channel ablations, using the multi-cycle curve fits to suppress single-cycle Monte Carlo statistical fluctuations), where stability accounts for $16\%$ of the combined logical error compared to $84\%$ from memory ($P^*_S / P^*_M \approx 0.19\times$ ratio in \fig[f]{2}; Table~\ref{tab:ablation_delta_ler_by_distance}). Fitting $P^*(d) = A\left(\mu\right)^{-(d+1)/2}$ across $d \in \{3, 5, 7\}$ yields the baseline inverse block suppression factors $1/\mu_{\mathrm{full}}$ and single-channel ablated values $1/\mu_{\mathrm{abl}, i}$. Table~\ref{tab:inverse_lambda_error_budget} reports the resulting channel sensitivities $\Delta(1/\mu_i) = 1/\mu_{\mathrm{full}} - 1/\mu_{\mathrm{abl}, i}$, normalized budget shares $\mathrm{Share}_i = \Delta(1/\mu_i) / \sum_j \Delta(1/\mu_j)$, and stability-to-memory relative contribution ratios $\mathrm{Share}_{T,i} / \mathrm{Share}_{S,i}$.

\begin{table}[t]
\centering
\small
\setlength{\tabcolsep}{5pt}
\caption{Space ($\mu_S$), Time ($\mu_T$), and Spacetime ($\mu_{\mathrm{ST}}$) block suppression error budgets fitted over odd distances ($d \in \{3, 5, 7\}$) at fixed optimal cycle count $c^*(d)$. Entries give sensitivity $\Delta(1/\mu)$ and relative budget share (\%). Propagated $1\sigma$ shot-noise uncertainties on $\Delta(1/\mu)$ are $\pm 0.002\text{--}0.004$ ($<1\%$ of budget) and are shown explicitly for the Stability/Memory ($T/S$) relative contribution ratio.}
\label{tab:inverse_lambda_error_budget}
\begin{tabular}{l rrr r}
\toprule
Channel & Memory ($S$) & Stability ($T$) & Spacetime ($ST$) & Ratio ($T/S$) \\
\midrule
2Q Gates & 0.336 (55.3\%) & 0.291 (43.4\%) & 0.327 (54.1\%) & $0.78\pm0.02\times$ \\
Idle & 0.193 (31.7\%) & 0.165 (24.5\%) & 0.189 (31.3\%) & $0.77\pm0.02\times$ \\
Readout & 0.030 (4.9\%) & 0.141 (21.0\%) & 0.037 (6.2\%) & $4.28\pm0.27\times$ \\
1Q Gates & 0.024 (4.0\%) & 0.028 (4.2\%) & 0.025 (4.1\%) & $1.05\pm0.12\times$ \\
Reset & 0.015 (2.5\%) & 0.041 (6.0\%) & 0.017 (2.7\%) & $2.43\pm0.34\times$ \\
DQLR & 0.010 (1.7\%) & 0.006 (0.9\%) & 0.009 (1.5\%) & $0.53\pm0.27\times$ \\
\midrule
Sum $\sum_i \Delta(1/\mu_i)$ & 0.608 (100\%) & 0.672 (100\%) & 0.605 (100\%) & --- \\
\bottomrule
\end{tabular}
\end{table}

\begin{table}[t]
\centering
\small
\setlength{\tabcolsep}{4.5pt}
\caption{Baseline block logical error probability $P^*_{\mathrm{full}}$ ($\times 10^{-3}$) and single-channel error reductions $\Delta P^* = P^*_{\mathrm{full}} - P^*_{\mathrm{abl}}$ ($\times 10^{-3}$) evaluated at the fixed optimal cycle count $c^*(d)$ across odd distances $d \in \{3, 5, 7\}$.}
\label{tab:ablation_delta_ler_by_distance}
\begin{tabular}{l c r rrrrrr}
\toprule
Metric & $d$ & Full & 2Q & Idle & RO & 1Q & Reset & DQLR \\
\midrule
Memory $P^*_M(c^*)$ & 3 ($c^*=4$) & 58.8 & 31.9 & 31.3 & 15.8 & 3.0 & 1.2 & 1.3 \\
 & 5 ($c^*=7$) & 43.3 & 33.3 & 29.9 & 12.1 & 3.6 & 2.2 & 1.6 \\
 & 7 ($c^*=9$) & 22.8 & 20.6 & 17.7 & 7.7 & 2.8 & 1.5 & 1.2 \\
\midrule
Stability $P^*_S(c^*)$ & 3 ($c^*=4$) & 14.5 & 8.3 & 4.3 & 8.4 & 1.0 & 3.7 & 0.1 \\
 & 5 ($c^*=7$) & 5.4 & 4.5 & 2.9 & 3.7 & 0.7 & 1.6 & 0.1 \\
 & 7 ($c^*=9$) & 4.7 & 4.2 & 3.0 & 3.6 & 0.7 & 1.7 & 0.1 \\
\midrule
Spacetime $P^*_{\mathrm{ST}}(c^*)$ & 3 ($c^*=4$) & 36.6 & 20.1 & 17.8 & 12.1 & 2.0 & 2.4 & 0.7 \\
 & 5 ($c^*=7$) & 24.3 & 18.9 & 16.4 & 7.9 & 2.2 & 1.9 & 0.8 \\
 & 7 ($c^*=9$) & 13.7 & 12.4 & 10.4 & 5.6 & 1.8 & 1.6 & 0.7 \\
\bottomrule
\end{tabular}
\end{table}

\section{Complementary gap} \label{sec:supp_complementary_gap}

\subsection{Decoding details} \label{sec:supp_decoding}

We decode the surface code stability experiments using a superconducting-inspired ($\mathrm{SI1000}$) circuit-level Detector Error Model (DEM) prior coupled with a correlated minimum-weight perfect matching (MWPM) decoder \cite{higgott_sparse_2023}. The $\mathrm{SI1000}$ noise model \cite{Gidney_2021_honeycomb} is parameterized by a single characteristic physical error probability $p=10^{-3}$, assigning two-qubit $\mathrm{CZ}$ gate depolarizing error $p$, single-qubit gate depolarizing error $p/10$, data-qubit idle error $p/10$ during gate layers, measurement error $5p$, reset bit-flip probability $2p$, and spectator data-qubit idle error $2p$ during measurement and reset windows.

\subsection{Definition and computation}
For a single experimental shot, let $W_{\mathrm{MLE}}$ be the total log-likelihood weight of the minimum-weight error configuration satisfying the syndrome constraints. Let $W_{\mathrm{comp}}(c)$ denote the weight of the minimum-weight error configuration consistent with the same syndrome but constrained to flip the logical observable to the opposite class. The \emph{complementary gap} is defined as the log-likelihood weight difference
\begin{equation}
  |G(c)| = W_{\mathrm{comp}}(c) - W_{\mathrm{MLE}}(c) \ge 0,
\end{equation}
and the \emph{signed complementary gap} relative to the true stability observable is
\begin{equation}
  G(c) =
  \begin{cases}
    +|G(c)|, & \text{if correctly decoded}, \\[4pt]
    -|G(c)|, & \text{if incorrectly decoded}.
  \end{cases}
\end{equation}
In our pipeline, $|G(c)|$ is computed at each candidate prefix window
$c \in [3, c_{\max}]$ ($c \in \{3, 4, \dots, c_{\max}\}$ for the static surface code,
with $c_{\max} = 60$ at $7 \times 7$ and $c_{\max} = 250$ at $3 \times 3$; and even
cycles $c \in \{4, 6, \dots, c_{\max}\}$ with $c_{\max} = 100$ for the two-cycle-period walking surface code) using a fixed logical boundary detector and two matching runs---once with the detector on and once with the detector off~\cite{gidney_yoked_2023}.

\subsection{Adaptive stopping rules}
In an online adaptive experiment, the controller observes only the unsigned gap $|G(c)|$ and the syndrome history up to cycle $c$, never the sign of $G(c)$. On most shots, errors are independent Pauli faults and $|G(c)|$ grows steadily in the correct logical class, allowing the experiment to terminate as soon as $|G(c)|$ crosses a confidence threshold $T$. During a multi-cycle leakage episode, however, the signed gap $G(c)$ can sometimes grow steadily negative---meaning the decoder becomes increasingly confident in the \emph{wrong} logical outcome. This happens because the decoder assumes all faults are independent: when a leaked qubit deposits a stream of detection events that locally pair up like natural, independent single-cycle errors, it can appear much cheaper to connect an initial defect to the early time boundary and treat the subsequent detections as independent local pairs than to track a single persistent fault across the entire window. Only after the leaked qubit relaxes back to the computational subspace does the true explanation become favored again, causing $G(c)$ to recover and grow positive.
Depending on whether the circuit architecture suppresses stationary multi-cycle leakage streaks, we employ two causal adaptive stopping strategies:
\begin{enumerate}
  \item \textbf{Pure gap stopping (used for the walking surface code):} A shot stops at the first candidate cycle $c$ where the complementary gap exceeds a fixed threshold,
  \begin{equation}
    |G(c)| \ge T.
  \end{equation}
  Because the walking surface code swaps data- and measure-qubit roles each cycle and applies echo gates, it is more difficult for a leaked qubit to sit on a fixed site and repeatedly deposit deceptive local detection pairs on the same checks; consequently, pure gap stopping suffices without an auxiliary syndrome filter, which can unnecessarily inflate the mean stopping time.
  \item \textbf{Syndrome-gated gap stopping (used for the standard surface code):} To prevent premature termination while a stationary leaked qubit is driving $G(c)$ negative, we pair the gap threshold with a causal local syndrome-burst indicator $F(u) \in \{0, 1\}$. At each detector round $u \le c - 1$, $F(u) = 1$ if any data qubit's trailing $W$-cycle neighborhood detection count exceeds a binomial $z$-score threshold $z^*$ above its baseline firing rate. A shot stops at the first cycle $c$ where the neighborhood is currently quiet and the gap exceeds a threshold raised by the cumulative number of previously flagged cycles $N_{\mathrm{flag}}(c) = \sum_{u=1}^{c-1} F(u)$:
  \begin{equation}
    F(c - 1) = 0 \qquad \text{and} \qquad |G(c)| \ge T + \beta\,N_{\mathrm{flag}}(c).
  \end{equation}
  Holding the shot while $F(c - 1) = 1$ prevents stopping mid-burst, while increasing the threshold by $\beta$ nats per flagged cycle ensures that once the burst ends, the experiment waits enough additional cycles for $G(c)$ to recover from negative values back above $+T$.
  In our analyses of the static $3 \times 3$ and $7 \times 7$ patches, each data qubit's neighborhood comprises its $m \le 4$ adjacent measure qubits tracked over a trailing window of $W = 7$ detector rounds, with baseline per-detector firing probability $p_0$ estimated from the empirical ensemble mean and burst threshold set to $z^* = 5$. The threshold-inflation slope is set to $\beta = 3.9\text{ nats}$ per flagged cycle for the $3 \times 3$ static patch (matching the mean Pauli gap growth rate) and $\beta = 3.1\text{ nats}$ per flagged cycle for the $7 \times 7$ patch; any shot not meeting the stopping criterion before the end of the recorded window terminates at $c_{\max}$. To generate the adaptive tradeoff curves in \fig[c,d]{5}, we sweep $T \in [2, 80]\text{ nats}$; for the $7 \times 7$ spacetime error minimum in \fig[d]{5} ($T^* = 8\text{ nats}$), the hyperparameters $(W, z^*, \beta, T^*)$ are selected in-sample on the hardware dataset, with out-of-sample generalization verified via a 50/50 train/test split across shots.
\end{enumerate}

\subsection{Enhanced adaptive gains under time-correlated errors}

While adaptive stopping already reduces the mean cycle count on Pauli-dominated shots by terminating high-confidence shots early, its advantage grows much larger in the presence of rarer, longer-lived time-correlated error mechanisms. The core intuition is simple: a fixed-cycle protocol must apply the same duration $c_0$ to every shot, forcing a tradeoff between stopping early on a correlated error floor or waiting long enough to clear rare bursts while accumulating extra spacelike memory error on every clean shot. Adaptive stopping breaks this tradeoff by finishing clean shots early and paying the long wait time only on the rare shots where a burst actually occurs.

\begin{figure}[]
  \centering
  \includegraphics[width=\textwidth]{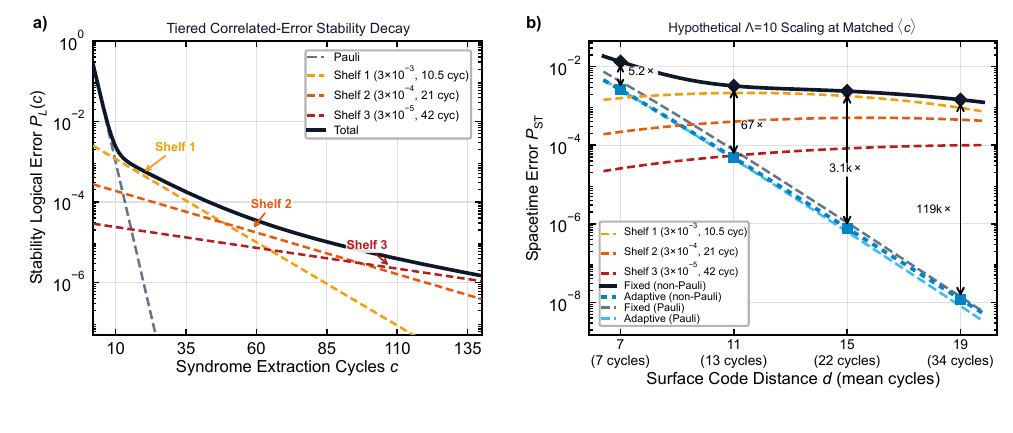}
  \caption{\textbf{Adaptive stopping bypasses tiered time-correlated error shelves in a toy model.}
\textbf{a}, Stability logical error $P_L(c)$ versus cycles $c$ at $d = 7$. The Pauli decay (gray dashed) and first correlated shelf ($3 \times 10^{-3}$, $10.5$ cycles; amber dashed) are matched to experiment, while two rarer hypothetical correlated mechanisms ($3 \times 10^{-4}$, $21$ cycles, orange dashed; $3 \times 10^{-5}$, $42$ cycles, crimson dashed) have $10\times$ smaller probability at each step and double in lifetime, producing a staircase of flatter error shelves in the total curve (solid black). \textbf{b}, Spacetime error $P_{\mathrm{ST}}$ versus code distance $d \in \{7, 11, 15, 19\}$ for a hypothetical $\Lambda = 10$, with optimal adaptive mean cycle count $\langle c \rangle_{\mathrm{adapt}}^* \approx \{7, 13, 22, 34\}$ cycles indicated on the horizontal axis. The Pauli-only adaptive lower bound (dashed light-blue) scales the optimal fixed-Pauli spacetime error at $c_{\mathrm{Pauli}}^*(d)$ (dashed gray) by the experimental early-stopping ratio at the same timelike Pauli error. Under a per-cycle patch burst rate $\gamma_k(d) = a_k(d)/\tau_k \propto d^2$, optimizing the adaptive stopping window to wait out rare bursts lifts the adaptive strategy with non-Pauli error (blue squares) by at most $1.5\times$ above the Pauli-only adaptive baseline. By contrast, a fixed-cycle strategy evaluated at the same mean cycle count $c_0(d) = \langle c \rangle_{\mathrm{adapt}}^*(d)$ (black diamonds) transitions from Pauli-limited at $d = 7$ to being pinned on the first residual shelf $a_1(d)e^{-c_0(d)/\tau_1}$ across $d = 11\text{--}19$ (with shelves~2 and~3 lying immediately below).
}
  \label{fig:toy_leakage_adaptive_advantage}
\end{figure}

To illustrate this effect in a toy model, we extend the two-component stability decay to include a hierarchy of correlated error tiers $k \in \{1, 2, 3\}$ with steady-state patch probability $a_k(d) \propto d^2$ and correlation lifetime $\tau_k$:
\begin{equation}
  P_L(c; d) = a_{\mathrm{Pauli}}(d)\,s_{\mathrm{Pauli}}(d)^{-c}
  + \sum_{k=1}^{3} a_k(d)\,e^{-c/\tau_k}.
\end{equation} 
Here tier~1 ($\tau_1 = 10.5$~cycles, $a_1(7) = 3.0 \times 10^{-3}$) matches the measured $\ket{3}$-leakage floor on the $7 \times 7$ patch, while hypothetical tiers $k = 2, 3$ occur $10\times$ less often at each step ($3.0 \times 10^{-4}$ and $3.0 \times 10^{-5}$ at $d = 7$) and persist twice as long ($\tau_2 = 21$ and $\tau_3 = 42$~cycles), producing a staircase of flatter error shelves (Fig.~\ref{fig:toy_leakage_adaptive_advantage}\textbf{a})---as could arise from excitations into progressively higher transmon states that cascade back down to the computational subspace.

To project spacetime scaling across $d \in \{7, 11, 15, 19\}$, we pair this stability decay with a hypothetical $\Lambda = 10$ per-cycle memory error rate $\epsilon_{\mathrm{mem}}(d) = \epsilon_{\mathrm{mem}}(7)\,\Lambda^{-(d-7)/2}$. Under adaptive stopping, burst-free shots terminate early at $c_{\mathrm{typ}}(d)$ cycles, while shots flagged on tier $k$ are extended by $\Delta c_k(d)$ cycles until the burst clears and its residual timelike error drops to the spacelike error scale:
\begin{equation}
  \langle c \rangle_{\mathrm{adapt}}(d) = c_{\mathrm{typ}}(d)
  + \sum_{k=1}^{3} p_{\mathrm{flag},k}(d)\,\Delta c_k(d).
\end{equation}
Because a tier-$k$ burst has steady-state patch probability $a_k(d)$ and lifetime $\tau_k$, bursts initiate across the patch at rate $\gamma_k(d) = a_k(d)/\tau_k \propto d^2$ per cycle. The probability $p_{\mathrm{flag},k}(d) \approx \gamma_k(d)\,c_{\mathrm{typ}}(d)$ of flagging a shot during the unflagged window therefore scales linearly with its spacetime volume $c_{\mathrm{typ}}(d) d^2$. Even at $d = 19$, over $95\%$ of shots finish unflagged, keeping the optimal mean runtime modest ($\langle c \rangle_{\mathrm{adapt}}^* \approx 7, 13, 22, 34$~cycles for $d = 7, 11, 15, 19$) and holding the adaptive spacetime error within $1.5\times$ of the Pauli-only adaptive baseline (Fig.~\ref{fig:toy_leakage_adaptive_advantage}\textbf{b}).

By contrast, a fixed-cycle strategy given the same average cycle budget $c_0(d) = \langle c \rangle_{\mathrm{adapt}}^*(d)$ never waits long enough ($\le 34$~cycles) to get past the first correlated shelf, stalling near $P_{\mathrm{ST}} \sim 10^{-3}$ across $d = 11\text{--}19$ (Fig.~\ref{fig:toy_leakage_adaptive_advantage}\textbf{b}). At matched mean runtime, adaptive stopping yields a spacetime error advantage that grows substantially from $d = 7$ to $d = 19$. Even if the fixed cycle count is instead globally increased to wait past the shelves (up to $\sim 400$~cycles at $d = 19$), adaptive stopping still achieves $2.1\times\text{--}12.5\times$ lower spacetime error while using $1.7\times\text{--}12\times$ fewer cycles on average.

\section{Walking Stability Experiment}
\label{sec:walking_stability}

\begin{figure*}[t]
    \centering
    \includegraphics[width=\textwidth]{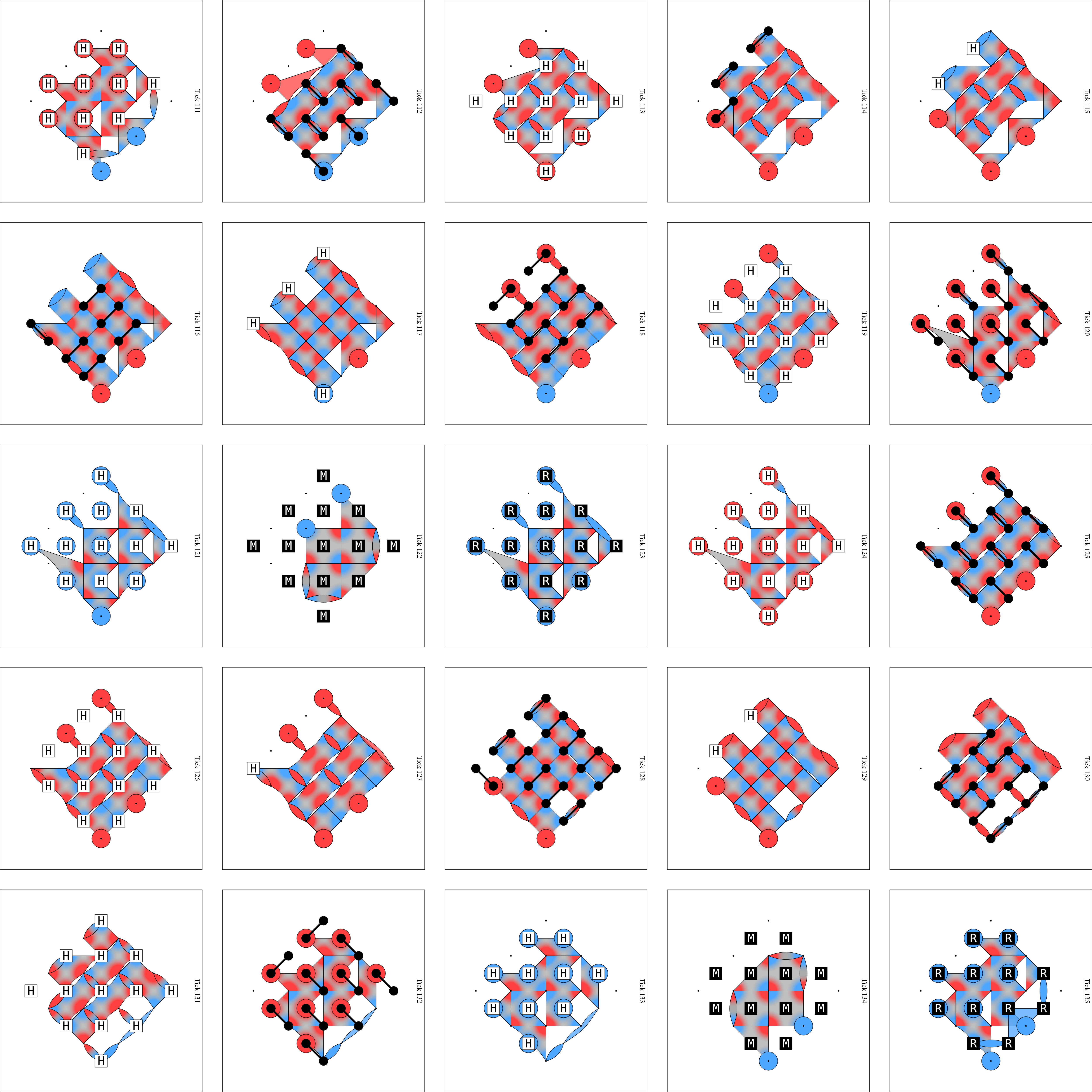}
    \caption{\textbf{Detector slices over two error-correction cycles of the $3\times 3$ walking stability experiment.}
    Stim detector-slice diagram (\texttt{detslice-with-ops}, read left-to-right, top-to-bottom) beginning and ending at a measurement layer (\texttt{M}) with two full syndrome-extraction cycles in between. Colored regions indicate the instantaneous Pauli support of each active detector ($X$ in red, $Y$ in green, $Z$ in blue). In the first cycle, the $3\times 3$ patch walks from the initial data sublattice onto the measure sublattice (middle measurement layer \texttt{M}); in the second cycle, it walks back to the original sublattice (final measurement layer \texttt{M}). At the measurement layers at the end of each cycle, weight-3 triangular boundary stabilizers pair with weight-1 single-qubit boundary detectors (blue circles on the boundary) to form effective weight-2 boundary stabilizers, preserving the $+1$ product of all boundary-type stabilizers.}
    \label{fig:walking_stability_detslices}
\end{figure*}

In a standard surface code stability experiment, physical qubits are permanently assigned as either data qubits or measure qubits. In the $3\times 3$ \emph{walking stability experiment}, the code patch instead translates back and forth across a $5\times 5$ grid of 25 physical qubits by swapping the roles of data and measure qubits in every cycle (Fig.~\ref{fig:walking_stability_detslices}). During each cycle's entangling layers, the stabilizers on the active data qubits are transferred onto the newly reset measure qubits before the old data qubits are measured and reset. Over two consecutive cycles, the patch walks by one lattice site from data qubits to measure qubits and then walks back to its original position.
A stability experiment requires that the product of all stabilizers of the boundary type equals $+1$, so that their combined syndrome outcomes form a deterministic global parity check in the absence of errors. In a static stability patch, this constraint is satisfied using weight-4 bulk stabilizers surrounded by weight-2 boundary stabilizers. In the walking circuit, as the patch shifts across the fixed qubit grid, the boundaries at the end of each cycle instead include weight-3 stabilizers paired with weight-1 single-qubit stabilizers/detectors on the outer boundary qubits (Fig.~\ref{fig:walking_stability_detslices}). Combining each weight-3 boundary stabilizer with its adjacent weight-1 boundary detector cancels the outer boundary qubit to form an effective weight-2 boundary stabilizer. Because every active data qubit is included an even number of times (twice) across the boundary-type stabilizers and detectors, their overall product remains $+1$ throughout the walk.

\begin{figure}[t]
    \centering
    \includegraphics[width=0.75\textwidth]{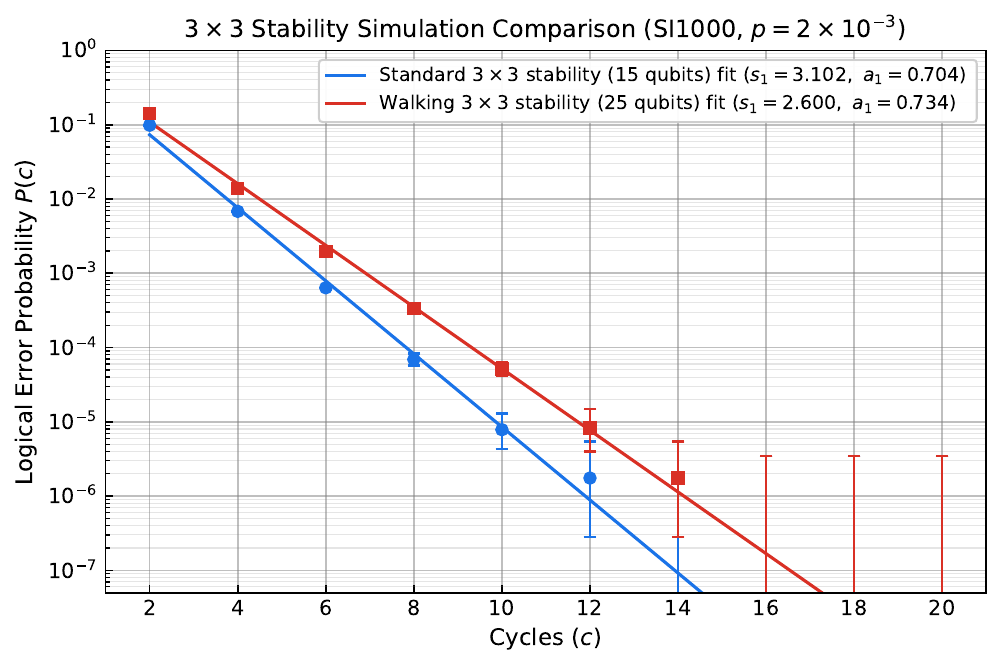}
    \caption{\textbf{Pauli model simulation of $3\times 3$ walking stability versus standard $3\times 3$ stability.}
    Logical error probability $P(c)$ as a function of the number of error-correction cycles $c$ (up to 20 cycles) under the SI1000 circuit-level noise model with $p = 2\times 10^{-3}$. Points are extracted via non-overlapping subsampling from 30-cycle Stim simulations ($2\times 10^6$ base shots) and decoded with Sparse Blossom. Error bars indicate binomial confidence intervals (\texttt{sinter}, likelihood ratio factor $10^3$; vertical bars without markers at $\ge 14$ cycles indicate zero-event upper bounds). Solid lines show exponential suppression fits $P(c) = a_1 (s_1)^{-c}$, yielding $s_1 = 3.102$ ($a_1 = 0.704$) for the standard $3\times 3$ stability circuit (15 physical qubits) and $s_1 = 2.600$ ($a_1 = 0.734$) for the walking $3\times 3$ stability circuit (25 physical qubits).}
    \label{fig:walking_vs_standard_stability_sim}
\end{figure}

To compare the baseline fault-tolerant performance of the walking circuit against a static patch, we simulate both the $3\times 3$ walking stability circuit and a standard $3\times 3$ stability circuit under the SI1000 circuit-level noise model at a physical error rate $p = 2\times 10^{-3}$ (Fig.~\ref{fig:walking_vs_standard_stability_sim}). Each circuit is simulated for 30 repeatable error-correction cycles ($2\times 10^6$ Monte Carlo shots) and analyzed using non-overlapping stability subsampling and minimum-weight perfect matching decoding. Both circuits exhibit exponential suppression of the logical error probability with cycle count $c$, $P(c) = a_1 (s_1)^{-c}$, with per-cycle suppression factors $s_1 = 3.10$ ($a_1 = 0.70$) for the standard $3\times 3$ stability circuit and $s_1 = 2.60$ ($a_1 = 0.73$) for the walking $3\times 3$ stability circuit. Because translating the patch across the grid and forming the weight-3 and weight-1 boundary detectors requires 25 physical qubits---compared to 15 physical qubits for the static $3\times 3$ stability circuit---the walking circuit incurs additional qubit and gate overhead, which accounts for the corresponding reduction in the error-suppression factor $s_1$.

\subsection*{Example 20-Cycle $3\times 3$ Walking Stability Stim Circuit}

This \href{https://algassert.com/crumble#circuit=Q(0,2)0;Q(0.5,1.5)1;Q(0.5,2.5)2;Q(1,1)3;Q(1,2)4;Q(1,3)5;Q(1.5,0.5)6;Q(1.5,1.5)7;Q(1.5,2.5)8;Q(1.5,3.5)9;Q(2,0)10;Q(2,1)11;Q(2,2)12;Q(2,3)13;Q(2,4)14;Q(2.5,0.5)15;Q(2.5,1.5)16;Q(2.5,2.5)17;Q(2.5,3.5)18;Q(3,1)19;Q(3,2)20;Q(3,3)21;Q(3.5,1.5)22;Q(3.5,2.5)23;Q(4,2)24;TICK;SQRT_X_9_18_23_17_8_7_16_22_15;TICK;H_24_23_21_20_19_18_16_14_13_12_11_10_8_5_4_3_0;TICK;CZ_23_20_22_19_18_13_17_12_16_11_15_10_9_5_8_4_7_3_2_0;TICK;H_23_22_18_17_16_15_8_7_2_0;TICK;CZ_24_23_22_20_21_18_19_16_17_13_15_11_12_8_10_6_7_4_3_1_2_0;TICK;H_24_19_10_9_6_3_1;TICK;CZ_23_21_20_17_18_14_16_12_13_9_11_7_10_6_8_5_4_2_3_1;TICK;H_21_20_14_13_12_6_5_1;TICK;CZ_24_22_21_17_20_16_19_15_14_9_13_8_12_7_5_2;TICK;H_22_17_16_15_9_8_7_2;TICK;M_17_2_7_15_9_22_1_6_23_18_16_8;OI(0)rec[-5]_rec[-6]_rec[-7]_rec[-8]_rec[-9]_rec[-10]_rec[-11]_rec[-12];TICK;R_15_7_9_17_22_1_2_6_23_18_16_8;TICK;H_22_17_16_15_9_8_7_6_2_1;TICK;CZ_24_22_21_17_20_16_19_15_14_9_13_8_12_7_5_2;TICK;H_21_20_14_13_12_5;TICK;CZ_23_21_20_17_18_14_16_12_13_9_11_7_10_6_8_5_4_2_3_1;TICK;H_24_19_14_10_9_6_3_1;TICK;CZ_24_23_22_20_21_18_19_16_17_13_15_11_12_8_10_6_7_4_3_1_2_0;TICK;H_24_23_22_21_18_17_16_15_8_7_2_0;TICK;CZ_23_20_22_19_18_13_17_12_16_11_15_10_9_5_8_4_7_3_2_0;TICK;H_20_19_13_12_11_10_5_4_3_0;TICK;M_5_14_12_21_19_24_3_10_0_13_20_4_11;DT(1.5,3.5,0)rec[-13]_rec[-21];DT(2.5,3.5,0)rec[-12]_rec[-16];DT(2.5,2.5,0)rec[-11]_rec[-25];DT(3.5,2.5,0)rec[-10]_rec[-16]_rec[-17];DT(3.5,1.5,0)rec[-8]_rec[-9]_rec[-17]_rec[-20];DT(0.5,1.5,0)rec[-7]_rec[-19]_rec[-23];DT(1.5,0.5,0)rec[-6]_rec[-18]_rec[-22];DT(0.5,2.5,0)rec[-5]_rec[-24];DT(1.5,2.5,0)rec[-2]_rec[-14];DT(2.5,1.5,0)rec[-1]_rec[-15];TICK;R_5_14_12_21_19_24_3_10_0_13_20_4_11;TICK;H_24_21_20_19_14_13_12_11_10_5_4_3_0;TICK;CZ_23_20_22_19_18_13_17_12_16_11_15_10_9_5_8_4_7_3_2_0;TICK;H_23_22_18_17_16_15_8_7_2_9_1_6;TICK;H_0;TICK;CZ_24_23_22_20_21_18_19_16_17_13_15_11_12_8_10_6_7_4_3_1_2_0;TICK;H_6_1;TICK;CZ_23_21_20_17_18_14_16_12_13_9_11_7_8_5_4_2;TICK;H_21_20_14_13_12_5_24_19_4_0_11_3_10;TICK;CZ_24_22_21_17_20_16_19_15_14_9_13_8_12_7_5_2_3_1_10_6;TICK;H_22_17_16_15_9_8_7_2_1_6;TICK;M_17_2_7_15_9_22_1_6_23_18_16_8;DT(3,3,1)rec[-12]_rec[-22];DT(0,2,1)rec[-11]_rec[-17]_rec[-25];DT(2,2,1)rec[-10]_rec[-23];DT(3,1,1)rec[-9]_rec[-21];DT(2,4,1)rec[-8]_rec[-24];DT(4,2,1)rec[-7]_rec[-20];DT(1,1,1)rec[-6]_rec[-19];DT(2,0,1)rec[-5]_rec[-18];DT(3,2,1)rec[-2]_rec[-15];DT(2,3,1)rec[-1]_rec[-16];TICK;R_15_7_9_17_22_1_2_6_23_18_16_8;TICK;H_22_17_16_15_9_8_7_6_2_1;TICK;CZ_24_22_21_17_20_16_19_15_14_9_13_8_12_7_5_2;TICK;H_21_20_14_13_12_5_24_19_4_11_0;TICK;CZ_10_6_3_1_2_4;TICK;H_1_6;TICK;CZ_20_17_16_12_13_9_11_7_8_5_2_0;TICK;H_14_10_3_0;TICK;CZ_22_20_19_16_17_13_15_11_12_8_10_6_7_4_3_1;TICK;H_23_22_18_17_16_15_8_7_2_9_1_6;TICK;CZ_23_20_22_19_18_13_17_12_16_11_15_10_9_5_8_4_7_3_2_0;TICK;H_20_19_13_12_11_10_5_4_3_0_24_21;TICK;M_5_14_12_21_19_24_3_10_0_13_20_4_11;DT(1.5,3.5,2)rec[-13]_rec[-21];DT(2.5,3.5,2)rec[-12]_rec[-16];DT(2.5,2.5,2)rec[-11]_rec[-25];DT(3.5,2.5,2)rec[-10]_rec[-16]_rec[-17];DT(3.5,1.5,2)rec[-8]_rec[-9]_rec[-17]_rec[-20];DT(0.5,1.5,2)rec[-7]_rec[-19]_rec[-23];DT(1.5,0.5,2)rec[-6]_rec[-18]_rec[-22];DT(0.5,2.5,2)rec[-5]_rec[-24];DT(1.5,2.5,2)rec[-2]_rec[-14];DT(2.5,1.5,2)rec[-1]_rec[-15];TICK;R_5_14_12_21_19_24_3_10_0_13_20_4_11;TICK;H_24_21_20_19_14_13_12_11_10_5_4_3_0;TICK;CZ_23_20_22_19_18_13_17_12_16_11_15_10_9_5_8_4_7_3_2_0;TICK;H_23_22_18_17_16_15_8_7_2_9_1_6;TICK;H_0;TICK;CZ_24_23_22_20_21_18_19_16_17_13_15_11_12_8_10_6_7_4_3_1_2_0;TICK;H_6_1;TICK;CZ_23_21_20_17_18_14_16_12_13_9_11_7_8_5_4_2;TICK;H_21_20_14_13_12_5_24_19_4_0_11_3_10;TICK;CZ_24_22_21_17_20_16_19_15_14_9_13_8_12_7_5_2_3_1_10_6;TICK;H_22_17_16_15_9_8_7_2_1_6;TICK;M_17_2_7_15_9_22_1_6_23_18_16_8;DT(3,3,3)rec[-12]_rec[-22];DT(0,2,3)rec[-11]_rec[-17]_rec[-25];DT(2,2,3)rec[-10]_rec[-23];DT(3,1,3)rec[-9]_rec[-21];DT(2,4,3)rec[-8]_rec[-24];DT(4,2,3)rec[-7]_rec[-20];DT(1,1,3)rec[-6]_rec[-19];DT(2,0,3)rec[-5]_rec[-18];DT(3,2,3)rec[-2]_rec[-15];DT(2,3,3)rec[-1]_rec[-16];TICK;R_15_7_9_17_22_1_2_6_23_18_16_8;TICK;H_22_17_16_15_9_8_7_6_2_1;TICK;CZ_24_22_21_17_20_16_19_15_14_9_13_8_12_7_5_2;TICK;H_21_20_14_13_12_5_24_19_4_11_0;TICK;CZ_10_6_3_1_2_4;TICK;H_1_6;TICK;CZ_20_17_16_12_13_9_11_7_8_5_2_0;TICK;H_14_10_3_0;TICK;CZ_22_20_19_16_17_13_15_11_12_8_10_6_7_4_3_1;TICK;H_23_22_18_17_16_15_8_7_2_9_1_6;TICK;CZ_23_20_22_19_18_13_17_12_16_11_15_10_9_5_8_4_7_3_2_0;TICK;H_20_19_13_12_11_10_5_4_3_0_24_21;TICK;M_5_14_12_21_19_24_3_10_0_13_20_4_11;DT(1.5,3.5,4)rec[-13]_rec[-21];DT(2.5,3.5,4)rec[-12]_rec[-16];DT(2.5,2.5,4)rec[-11]_rec[-25];DT(3.5,2.5,4)rec[-10]_rec[-16]_rec[-17];DT(3.5,1.5,4)rec[-8]_rec[-9]_rec[-17]_rec[-20];DT(0.5,1.5,4)rec[-7]_rec[-19]_rec[-23];DT(1.5,0.5,4)rec[-6]_rec[-18]_rec[-22];DT(0.5,2.5,4)rec[-5]_rec[-24];DT(1.5,2.5,4)rec[-2]_rec[-14];DT(2.5,1.5,4)rec[-1]_rec[-15];TICK;R_5_14_12_21_19_24_3_10_0_13_20_4_11;TICK;H_24_21_20_19_14_13_12_11_10_5_4_3_0;TICK;CZ_23_20_22_19_18_13_17_12_16_11_15_10_9_5_8_4_7_3_2_0;TICK;H_23_22_18_17_16_15_8_7_2_9_1_6;TICK;H_0;TICK;CZ_24_23_22_20_21_18_19_16_17_13_15_11_12_8_10_6_7_4_3_1_2_0;TICK;H_6_1;TICK;CZ_23_21_20_17_18_14_16_12_13_9_11_7_8_5_4_2;TICK;H_21_20_14_13_12_5_24_19_4_0_11_3_10;TICK;CZ_24_22_21_17_20_16_19_15_14_9_13_8_12_7_5_2_3_1_10_6;TICK;H_22_17_16_15_9_8_7_2_1_6;TICK;M_17_2_7_15_9_22_1_6_23_18_16_8;DT(3,3,5)rec[-12]_rec[-22];DT(0,2,5)rec[-11]_rec[-17]_rec[-25];DT(2,2,5)rec[-10]_rec[-23];DT(3,1,5)rec[-9]_rec[-21];DT(2,4,5)rec[-8]_rec[-24];DT(4,2,5)rec[-7]_rec[-20];DT(1,1,5)rec[-6]_rec[-19];DT(2,0,5)rec[-5]_rec[-18];DT(3,2,5)rec[-2]_rec[-15];DT(2,3,5)rec[-1]_rec[-16];TICK;R_15_7_9_17_22_1_2_6_23_18_16_8;TICK;H_22_17_16_15_9_8_7_6_2_1;TICK;CZ_24_22_21_17_20_16_19_15_14_9_13_8_12_7_5_2;TICK;H_21_20_14_13_12_5_24_19_4_11_0;TICK;CZ_10_6_3_1_2_4;TICK;H_1_6;TICK;CZ_20_17_16_12_13_9_11_7_8_5_2_0;TICK;H_14_10_3_0;TICK;CZ_22_20_19_16_17_13_15_11_12_8_10_6_7_4_3_1;TICK;H_23_22_18_17_16_15_8_7_2_9_1_6;TICK;CZ_23_20_22_19_18_13_17_12_16_11_15_10_9_5_8_4_7_3_2_0;TICK;H_20_19_13_12_11_10_5_4_3_0_24_21;TICK;M_5_14_12_21_19_24_3_10_0_13_20_4_11;DT(1.5,3.5,6)rec[-13]_rec[-21];DT(2.5,3.5,6)rec[-12]_rec[-16];DT(2.5,2.5,6)rec[-11]_rec[-25];DT(3.5,2.5,6)rec[-10]_rec[-16]_rec[-17];DT(3.5,1.5,6)rec[-8]_rec[-9]_rec[-17]_rec[-20];DT(0.5,1.5,6)rec[-7]_rec[-19]_rec[-23];DT(1.5,0.5,6)rec[-6]_rec[-18]_rec[-22];DT(0.5,2.5,6)rec[-5]_rec[-24];DT(1.5,2.5,6)rec[-2]_rec[-14];DT(2.5,1.5,6)rec[-1]_rec[-15];TICK;R_5_14_12_21_19_24_3_10_0_13_20_4_11;TICK;H_24_21_20_19_14_13_12_11_10_5_4_3_0;TICK;CZ_23_20_22_19_18_13_17_12_16_11_15_10_9_5_8_4_7_3_2_0;TICK;H_23_22_18_17_16_15_8_7_2_9_1_6;TICK;H_0;TICK;CZ_24_23_22_20_21_18_19_16_17_13_15_11_12_8_10_6_7_4_3_1_2_0;TICK;H_6_1;TICK;CZ_23_21_20_17_18_14_16_12_13_9_11_7_8_5_4_2;TICK;H_21_20_14_13_12_5_24_19_4_0_11_3_10;TICK;CZ_24_22_21_17_20_16_19_15_14_9_13_8_12_7_5_2_3_1_10_6;TICK;H_22_17_16_15_9_8_7_2_1_6;TICK;M_17_2_7_15_9_22_1_6_23_18_16_8;DT(3,3,7)rec[-12]_rec[-22];DT(0,2,7)rec[-11]_rec[-17]_rec[-25];DT(2,2,7)rec[-10]_rec[-23];DT(3,1,7)rec[-9]_rec[-21];DT(2,4,7)rec[-8]_rec[-24];DT(4,2,7)rec[-7]_rec[-20];DT(1,1,7)rec[-6]_rec[-19];DT(2,0,7)rec[-5]_rec[-18];DT(3,2,7)rec[-2]_rec[-15];DT(2,3,7)rec[-1]_rec[-16];TICK;R_15_7_9_17_22_1_2_6_23_18_16_8;TICK;H_22_17_16_15_9_8_7_6_2_1;TICK;CZ_24_22_21_17_20_16_19_15_14_9_13_8_12_7_5_2;TICK;H_21_20_14_13_12_5_24_19_4_11_0;TICK;CZ_10_6_3_1_2_4;TICK;H_1_6;TICK;CZ_20_17_16_12_13_9_11_7_8_5_2_0;TICK;H_14_10_3_0;TICK;CZ_22_20_19_16_17_13_15_11_12_8_10_6_7_4_3_1;TICK;H_23_22_18_17_16_15_8_7_2_9_1_6;TICK;CZ_23_20_22_19_18_13_17_12_16_11_15_10_9_5_8_4_7_3_2_0;TICK;H_20_19_13_12_11_10_5_4_3_0_24_21;TICK;M_5_14_12_21_19_24_3_10_0_13_20_4_11;DT(1.5,3.5,8)rec[-13]_rec[-21];DT(2.5,3.5,8)rec[-12]_rec[-16];DT(2.5,2.5,8)rec[-11]_rec[-25];DT(3.5,2.5,8)rec[-10]_rec[-16]_rec[-17];DT(3.5,1.5,8)rec[-8]_rec[-9]_rec[-17]_rec[-20];DT(0.5,1.5,8)rec[-7]_rec[-19]_rec[-23];DT(1.5,0.5,8)rec[-6]_rec[-18]_rec[-22];DT(0.5,2.5,8)rec[-5]_rec[-24];DT(1.5,2.5,8)rec[-2]_rec[-14];DT(2.5,1.5,8)rec[-1]_rec[-15];TICK;R_5_14_12_21_19_24_3_10_0_13_20_4_11;TICK;H_24_21_20_19_14_13_12_11_10_5_4_3_0;TICK;CZ_23_20_22_19_18_13_17_12_16_11_15_10_9_5_8_4_7_3_2_0;TICK;H_23_22_18_17_16_15_8_7_2_9_1_6;TICK;H_0;TICK;CZ_24_23_22_20_21_18_19_16_17_13_15_11_12_8_10_6_7_4_3_1_2_0;TICK;H_6_1;TICK;CZ_23_21_20_17_18_14_16_12_13_9_11_7_8_5_4_2;TICK;H_21_20_14_13_12_5_24_19_4_0_11_3_10;TICK;CZ_24_22_21_17_20_16_19_15_14_9_13_8_12_7_5_2_3_1_10_6;TICK;H_22_17_16_15_9_8_7_2_1_6;TICK;M_17_2_7_15_9_22_1_6_23_18_16_8;DT(3,3,9)rec[-12]_rec[-22];DT(0,2,9)rec[-11]_rec[-17]_rec[-25];DT(2,2,9)rec[-10]_rec[-23];DT(3,1,9)rec[-9]_rec[-21];DT(2,4,9)rec[-8]_rec[-24];DT(4,2,9)rec[-7]_rec[-20];DT(1,1,9)rec[-6]_rec[-19];DT(2,0,9)rec[-5]_rec[-18];DT(3,2,9)rec[-2]_rec[-15];DT(2,3,9)rec[-1]_rec[-16];TICK;R_15_7_9_17_22_1_2_6_23_18_16_8;TICK;H_22_17_16_15_9_8_7_6_2_1;TICK;CZ_24_22_21_17_20_16_19_15_14_9_13_8_12_7_5_2;TICK;H_21_20_14_13_12_5_24_19_4_11_0;TICK;CZ_10_6_3_1_2_4;TICK;H_1_6;TICK;CZ_20_17_16_12_13_9_11_7_8_5_2_0;TICK;H_14_10_3_0;TICK;CZ_22_20_19_16_17_13_15_11_12_8_10_6_7_4_3_1;TICK;H_23_22_18_17_16_15_8_7_2_9_1_6;TICK;CZ_23_20_22_19_18_13_17_12_16_11_15_10_9_5_8_4_7_3_2_0;TICK;H_20_19_13_12_11_10_5_4_3_0_24_21;TICK;M_5_14_12_21_19_24_3_10_0_13_20_4_11;DT(1.5,3.5,10)rec[-13]_rec[-21];DT(2.5,3.5,10)rec[-12]_rec[-16];DT(2.5,2.5,10)rec[-11]_rec[-25];DT(3.5,2.5,10)rec[-10]_rec[-16]_rec[-17];DT(3.5,1.5,10)rec[-8]_rec[-9]_rec[-17]_rec[-20];DT(0.5,1.5,10)rec[-7]_rec[-19]_rec[-23];DT(1.5,0.5,10)rec[-6]_rec[-18]_rec[-22];DT(0.5,2.5,10)rec[-5]_rec[-24];DT(1.5,2.5,10)rec[-2]_rec[-14];DT(2.5,1.5,10)rec[-1]_rec[-15];TICK;R_5_14_12_21_19_24_3_10_0_13_20_4_11;TICK;H_24_21_20_19_14_13_12_11_10_5_4_3_0;TICK;CZ_23_20_22_19_18_13_17_12_16_11_15_10_9_5_8_4_7_3_2_0;TICK;H_23_22_18_17_16_15_8_7_2_9_1_6;TICK;H_0;TICK;CZ_24_23_22_20_21_18_19_16_17_13_15_11_12_8_10_6_7_4_3_1_2_0;TICK;H_6_1;TICK;CZ_23_21_20_17_18_14_16_12_13_9_11_7_8_5_4_2;TICK;H_21_20_14_13_12_5_24_19_4_0_11_3_10;TICK;CZ_24_22_21_17_20_16_19_15_14_9_13_8_12_7_5_2_3_1_10_6;TICK;H_22_17_16_15_9_8_7_2_1_6;TICK;M_17_2_7_15_9_22_1_6_23_18_16_8;DT(3,3,11)rec[-12]_rec[-22];DT(0,2,11)rec[-11]_rec[-17]_rec[-25];DT(2,2,11)rec[-10]_rec[-23];DT(3,1,11)rec[-9]_rec[-21];DT(2,4,11)rec[-8]_rec[-24];DT(4,2,11)rec[-7]_rec[-20];DT(1,1,11)rec[-6]_rec[-19];DT(2,0,11)rec[-5]_rec[-18];DT(3,2,11)rec[-2]_rec[-15];DT(2,3,11)rec[-1]_rec[-16];TICK;R_15_7_9_17_22_1_2_6_23_18_16_8;TICK;H_22_17_16_15_9_8_7_6_2_1;TICK;CZ_24_22_21_17_20_16_19_15_14_9_13_8_12_7_5_2;TICK;H_21_20_14_13_12_5_24_19_4_11_0;TICK;CZ_10_6_3_1_2_4;TICK;H_1_6;TICK;CZ_20_17_16_12_13_9_11_7_8_5_2_0;TICK;H_14_10_3_0;TICK;CZ_22_20_19_16_17_13_15_11_12_8_10_6_7_4_3_1;TICK;H_23_22_18_17_16_15_8_7_2_9_1_6;TICK;CZ_23_20_22_19_18_13_17_12_16_11_15_10_9_5_8_4_7_3_2_0;TICK;H_20_19_13_12_11_10_5_4_3_0_24_21;TICK;M_5_14_12_21_19_24_3_10_0_13_20_4_11;DT(1.5,3.5,12)rec[-13]_rec[-21];DT(2.5,3.5,12)rec[-12]_rec[-16];DT(2.5,2.5,12)rec[-11]_rec[-25];DT(3.5,2.5,12)rec[-10]_rec[-16]_rec[-17];DT(3.5,1.5,12)rec[-8]_rec[-9]_rec[-17]_rec[-20];DT(0.5,1.5,12)rec[-7]_rec[-19]_rec[-23];DT(1.5,0.5,12)rec[-6]_rec[-18]_rec[-22];DT(0.5,2.5,12)rec[-5]_rec[-24];DT(1.5,2.5,12)rec[-2]_rec[-14];DT(2.5,1.5,12)rec[-1]_rec[-15];TICK;R_5_14_12_21_19_24_3_10_0_13_20_4_11;TICK;H_24_21_20_19_14_13_12_11_10_5_4_3_0;TICK;CZ_23_20_22_19_18_13_17_12_16_11_15_10_9_5_8_4_7_3_2_0;TICK;H_23_22_18_17_16_15_8_7_2_9_1_6;TICK;H_0;TICK;CZ_24_23_22_20_21_18_19_16_17_13_15_11_12_8_10_6_7_4_3_1_2_0;TICK;H_6_1;TICK;CZ_23_21_20_17_18_14_16_12_13_9_11_7_8_5_4_2;TICK;H_21_20_14_13_12_5_24_19_4_0_11_3_10;TICK;CZ_24_22_21_17_20_16_19_15_14_9_13_8_12_7_5_2_3_1_10_6;TICK;H_22_17_16_15_9_8_7_2_1_6;TICK;M_17_2_7_15_9_22_1_6_23_18_16_8;DT(3,3,13)rec[-12]_rec[-22];DT(0,2,13)rec[-11]_rec[-17]_rec[-25];DT(2,2,13)rec[-10]_rec[-23];DT(3,1,13)rec[-9]_rec[-21];DT(2,4,13)rec[-8]_rec[-24];DT(4,2,13)rec[-7]_rec[-20];DT(1,1,13)rec[-6]_rec[-19];DT(2,0,13)rec[-5]_rec[-18];DT(3,2,13)rec[-2]_rec[-15];DT(2,3,13)rec[-1]_rec[-16];TICK;R_15_7_9_17_22_1_2_6_23_18_16_8;TICK;H_22_17_16_15_9_8_7_6_2_1;TICK;CZ_24_22_21_17_20_16_19_15_14_9_13_8_12_7_5_2;TICK;H_21_20_14_13_12_5_24_19_4_11_0;TICK;CZ_10_6_3_1_2_4;TICK;H_1_6;TICK;CZ_20_17_16_12_13_9_11_7_8_5_2_0;TICK;H_14_10_3_0;TICK;CZ_22_20_19_16_17_13_15_11_12_8_10_6_7_4_3_1;TICK;H_23_22_18_17_16_15_8_7_2_9_1_6;TICK;CZ_23_20_22_19_18_13_17_12_16_11_15_10_9_5_8_4_7_3_2_0;TICK;H_20_19_13_12_11_10_5_4_3_0_24_21;TICK;M_5_14_12_21_19_24_3_10_0_13_20_4_11;DT(1.5,3.5,14)rec[-13]_rec[-21];DT(2.5,3.5,14)rec[-12]_rec[-16];DT(2.5,2.5,14)rec[-11]_rec[-25];DT(3.5,2.5,14)rec[-10]_rec[-16]_rec[-17];DT(3.5,1.5,14)rec[-8]_rec[-9]_rec[-17]_rec[-20];DT(0.5,1.5,14)rec[-7]_rec[-19]_rec[-23];DT(1.5,0.5,14)rec[-6]_rec[-18]_rec[-22];DT(0.5,2.5,14)rec[-5]_rec[-24];DT(1.5,2.5,14)rec[-2]_rec[-14];DT(2.5,1.5,14)rec[-1]_rec[-15];TICK;R_5_14_12_21_19_24_3_10_0_13_20_4_11;TICK;H_24_21_20_19_14_13_12_11_10_5_4_3_0;TICK;CZ_23_20_22_19_18_13_17_12_16_11_15_10_9_5_8_4_7_3_2_0;TICK;H_23_22_18_17_16_15_8_7_2_9_1_6;TICK;H_0;TICK;CZ_24_23_22_20_21_18_19_16_17_13_15_11_12_8_10_6_7_4_3_1_2_0;TICK;H_6_1;TICK;CZ_23_21_20_17_18_14_16_12_13_9_11_7_8_5_4_2;TICK;H_21_20_14_13_12_5_24_19_4_0_11_3_10;TICK;CZ_24_22_21_17_20_16_19_15_14_9_13_8_12_7_5_2_3_1_10_6;TICK;H_22_17_16_15_9_8_7_2_1_6;TICK;M_17_2_7_15_9_22_1_6_23_18_16_8;DT(3,3,15)rec[-12]_rec[-22];DT(0,2,15)rec[-11]_rec[-17]_rec[-25];DT(2,2,15)rec[-10]_rec[-23];DT(3,1,15)rec[-9]_rec[-21];DT(2,4,15)rec[-8]_rec[-24];DT(4,2,15)rec[-7]_rec[-20];DT(1,1,15)rec[-6]_rec[-19];DT(2,0,15)rec[-5]_rec[-18];DT(3,2,15)rec[-2]_rec[-15];DT(2,3,15)rec[-1]_rec[-16];TICK;R_15_7_9_17_22_1_2_6_23_18_16_8;TICK;H_22_17_16_15_9_8_7_6_2_1;TICK;CZ_24_22_21_17_20_16_19_15_14_9_13_8_12_7_5_2;TICK;H_21_20_14_13_12_5_24_19_4_11_0;TICK;CZ_10_6_3_1_2_4;TICK;H_1_6;TICK;CZ_20_17_16_12_13_9_11_7_8_5_2_0;TICK;H_14_10_3_0;TICK;CZ_22_20_19_16_17_13_15_11_12_8_10_6_7_4_3_1;TICK;H_23_22_18_17_16_15_8_7_2_9_1_6;TICK;CZ_23_20_22_19_18_13_17_12_16_11_15_10_9_5_8_4_7_3_2_0;TICK;H_20_19_13_12_11_10_5_4_3_0_24_21;TICK;M_5_14_12_21_19_24_3_10_0_13_20_4_11;DT(1.5,3.5,16)rec[-13]_rec[-21];DT(2.5,3.5,16)rec[-12]_rec[-16];DT(2.5,2.5,16)rec[-11]_rec[-25];DT(3.5,2.5,16)rec[-10]_rec[-16]_rec[-17];DT(3.5,1.5,16)rec[-8]_rec[-9]_rec[-17]_rec[-20];DT(0.5,1.5,16)rec[-7]_rec[-19]_rec[-23];DT(1.5,0.5,16)rec[-6]_rec[-18]_rec[-22];DT(0.5,2.5,16)rec[-5]_rec[-24];DT(1.5,2.5,16)rec[-2]_rec[-14];DT(2.5,1.5,16)rec[-1]_rec[-15];TICK;R_5_14_12_21_19_24_3_10_0_13_20_4_11;TICK;H_24_21_20_19_14_13_12_11_10_5_4_3_0;TICK;CZ_23_20_22_19_18_13_17_12_16_11_15_10_9_5_8_4_7_3_2_0;TICK;H_23_22_18_17_16_15_8_7_2_0;TICK;CZ_24_23_22_20_21_18_19_16_17_13_15_11_12_8_10_6_7_4_3_1_2_0;TICK;H_24_19_10_9_6_3_1;TICK;CZ_23_21_20_17_18_14_16_12_13_9_11_7_10_6_8_5_4_2_3_1;TICK;H_21_20_14_13_12_6_5_1;TICK;CZ_24_22_21_17_20_16_19_15_14_9_13_8_12_7_5_2;TICK;H_22_17_16_15_9_8_7_2;TICK;M_17_2_7_15_9_22_1_6_23_18_16_8;DT(3,3,17)rec[-12]_rec[-22];DT(0,2,17)rec[-11]_rec[-17]_rec[-25];DT(2,2,17)rec[-10]_rec[-23];DT(3,1,17)rec[-9]_rec[-21];DT(2,4,17)rec[-8]_rec[-24];DT(4,2,17)rec[-7]_rec[-20];DT(1,1,17)rec[-6]_rec[-19];DT(2,0,17)rec[-5]_rec[-18];DT(3,2,17)rec[-2]_rec[-15];DT(2,3,17)rec[-1]_rec[-16];TICK;R_15_7_9_17_22_1_2_6_23_18_16_8;TICK;H_22_17_16_15_9_8_7_6_2_1;TICK;CZ_24_22_21_17_20_16_19_15_14_9_13_8_12_7_5_2;TICK;H_21_20_14_13_12_5;TICK;CZ_23_21_20_17_18_14_16_12_13_9_11_7_10_6_8_5_4_2_3_1;TICK;H_24_19_14_10_9_6_3_1;TICK;CZ_24_23_22_20_21_18_19_16_17_13_15_11_12_8_10_6_7_4_3_1_2_0;TICK;H_24_23_22_21_18_17_16_15_8_7_2_0;TICK;CZ_23_20_22_19_18_13_17_12_16_11_15_10_9_5_8_4_7_3_2_0;TICK;H_20_19_13_12_11_10_5_4_3_0;SQRT_X_9_18_23_17_8_7_16_22_15;TICK;M_5_14_12_21_19_24_3_10_0_13_20_4_11_9_18_23_17_8_7_16_22_15;DT(1.5,3.5,18)rec[-22]_rec[-30];DT(2.5,3.5,18)rec[-21]_rec[-25];DT(2.5,2.5,18)rec[-20]_rec[-34];DT(3.5,2.5,18)rec[-19]_rec[-25]_rec[-26];DT(3.5,1.5,18)rec[-17]_rec[-18]_rec[-26]_rec[-29];DT(0.5,1.5,18)rec[-16]_rec[-28]_rec[-32];DT(1.5,0.5,18)rec[-15]_rec[-27]_rec[-31];DT(0.5,2.5,18)rec[-14]_rec[-33];DT(1.5,2.5,18)rec[-11]_rec[-23];DT(2.5,1.5,18)rec[-10]_rec[-24]}{crumble link} is the complete Stim circuit for a 20-cycle $3\times 3$ walking stability experiment (2 initialization cycles, 7 repetitions of the 2-cycle bulk loop, and 4 final cycles).

\section{Stability Experiment Subsampling}
\label{sec:supp_stability_subsampling}

\begin{figure}[t]
    \centering
    \includegraphics[width=0.7\textwidth]{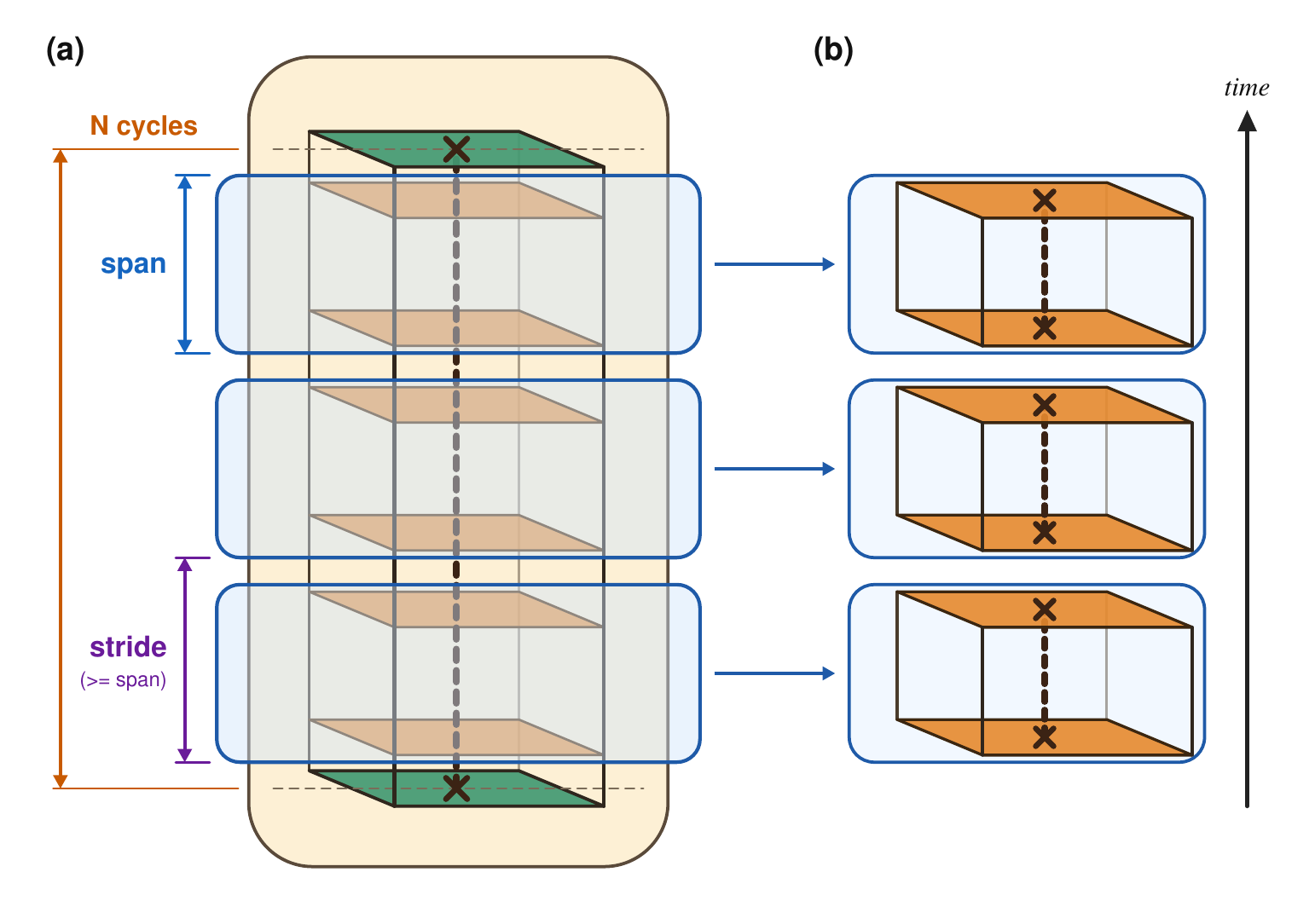}
    \caption{\textbf{Subsampling a stability experiment.}
    \textbf{(a)}~Spacetime volume of an $N$-cycle stability experiment (beige), bounded by initial and final spacelike stabilizer planes (green) and a timelike stability observable (vertical dashed line). The bulk measurement record is sliced into windows (light blue) of duration $\text{span}$ cycles separated by $\text{stride}$ cycles.
    \textbf{(b)}~Each window is decoded independently as a $\text{span}$-cycle stability experiment bounded by its first and last syndrome layers (orange planes).}
    \label{fig:stability_subsampling}
\end{figure}

The bulk of an $N$-cycle stability experiment consists of identical, repeated syndrome-extraction cycles.
As such, a single long run naturally contains the measurement history of shorter stability experiments within its interior. Rather than executing separate experiments at every cycle duration, we use \emph{subsampling} to extract multiple shorter stability experiments from one long run (Fig.~\ref{fig:stability_subsampling}).

In post-processing, we slice the bulk measurement record into $M$ smaller windows separated by a start-to-start offset $\text{stride}$ (Fig.~\ref{fig:stability_subsampling}a).
The first and last syndrome layers of each window define its temporal boundaries (Fig.~\ref{fig:stability_subsampling}b); detectors referencing earlier measurements are dropped, and each window is decoded independently.
Sweeping window size across the same dataset yields the full logical error probability versus cycle curve from a single experiment.

Extracting $M$ windows from $N_{\text{shots}}$ repetitions gives $M \times N_{\text{shots}}$ subsampled shots. For non-overlapping windows ($\text{stride} \ge \text{span}$), all $N_{\text{eff}} = M \times N_{\text{shots}}$ samples are independent. When windows partially overlap ($\text{stride} < \text{span}$), shared measurements introduce correlations; we account for this by scaling down the raw shot count by the fraction of unique cycles covered (total distinct cycles spanned across all windows divided by $M \times \text{span}$) and computing binomial error bars from the resulting effective shot count $N_{\text{eff}}$.
For the walking experiment data in \fig[a]{4}, all windows are independent.

\begin{figure*}[t]
    \centering
    \includegraphics[width=\textwidth]{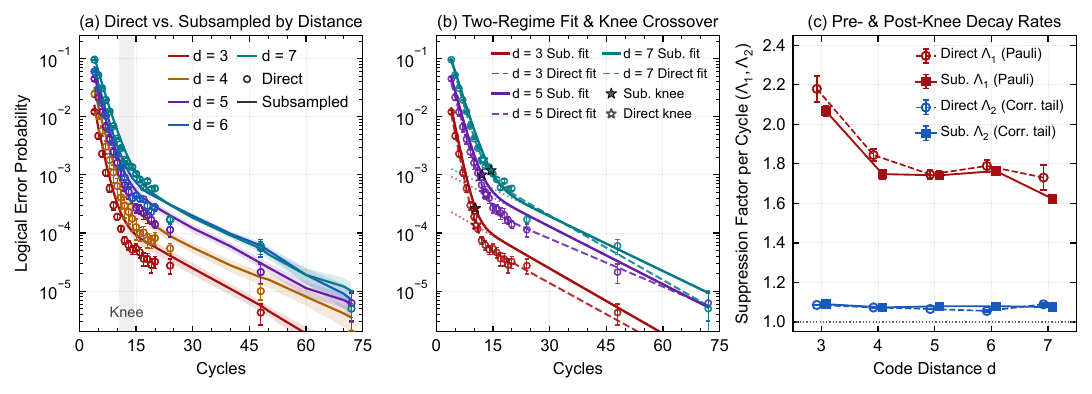}
    \caption{\textbf{Experimental validation of stability subsampling across code distances $d = 3\text{--}7$.}
    \textbf{(a)}~Patch-averaged logical error probability versus cycles $c$ for direct runs (open circles) and subsampled windows (solid curves, $\pm 1\sigma$ bands from $N_{\text{eff}}$) from the $N = 240$-cycle run. Subsampling slightly over-predicts the error rate while accurately capturing the knee (gray band) and post-knee tail.
    \textbf{(b)}~Two-exponential fits for $d = 3, 5, 7$ comparing subsampled (solid) and direct (dashed) curves, with correlated-error components $a_2 s_2^{-c}$ (dotted) and knee crossovers (stars).
    \textbf{(c)}~Fitted pre-knee ($s_1$, red) and post-knee ($s_2$, blue) suppression factors versus code distance $d$ for direct (open circles) and subsampled (filled squares) data.}
    \label{fig:stability_subsampling_validation}
\end{figure*}

We validate subsampling against direct stability experiments swept across cycle counts $c$ on $15$ surface code patches ($d = 3\text{--}7$). Subsampled curves are extracted solely from the $N = 240$-cycle run of each patch and decoded with the same settings as the direct runs.
Error bars and shaded bands in Fig.~\ref{fig:stability_subsampling_validation}a combine patch-to-patch standard errors with the overlap-scaled binomial uncertainty from $N_{\text{eff}}$.

Across all five code distances, subsampling closely tracks the direct experiments with a small upward offset (Fig.~\ref{fig:stability_subsampling_validation}a). Two effects make the subsampled error rate slightly higher: (1)~\emph{temporal edge effects}, because subsampled windows terminate on truncated bulk syndrome layers (orange planes in Fig.~\ref{fig:stability_subsampling}b) rather than physical state preparation and destructive data-qubit readout (green planes in Fig.~\ref{fig:stability_subsampling}a); and (2)~\emph{steady-state leakage}, which ramps up over the first handful of cycles---direct runs always start from a freshly reset $\ket{0}$ state on all data and measure qubits, whereas subsampled windows are drawn from the interior of a $240$-cycle run after leakage has equilibrated. Subsampling thus provides a slightly conservative upper bound on performance.

Crucially, subsampling accurately captures both the \emph{knee} and the \emph{correlated-error regime} beyond it. Using the two-exponential fit from the main text (Fig.~\ref{fig:stability_subsampling_validation}b), subsampling reproduces the knee location ($c \approx 11\text{--}14$) to within $1\text{--}2$ cycles and recovers both the pre-knee Pauli suppression factor $s_1$ and the post-knee correlated-error suppression factor $s_2$ across $d = 3\text{--}7$ (agreeing within $\sim 2\%$ for $s_2$; Fig.~\ref{fig:stability_subsampling_validation}c), while reducing statistical scatter in the low-error tail.
\section{Reinforcement learning calibration} \label{sec:supp_rl_calibration}

\begin{figure}
    \centering
    \includegraphics{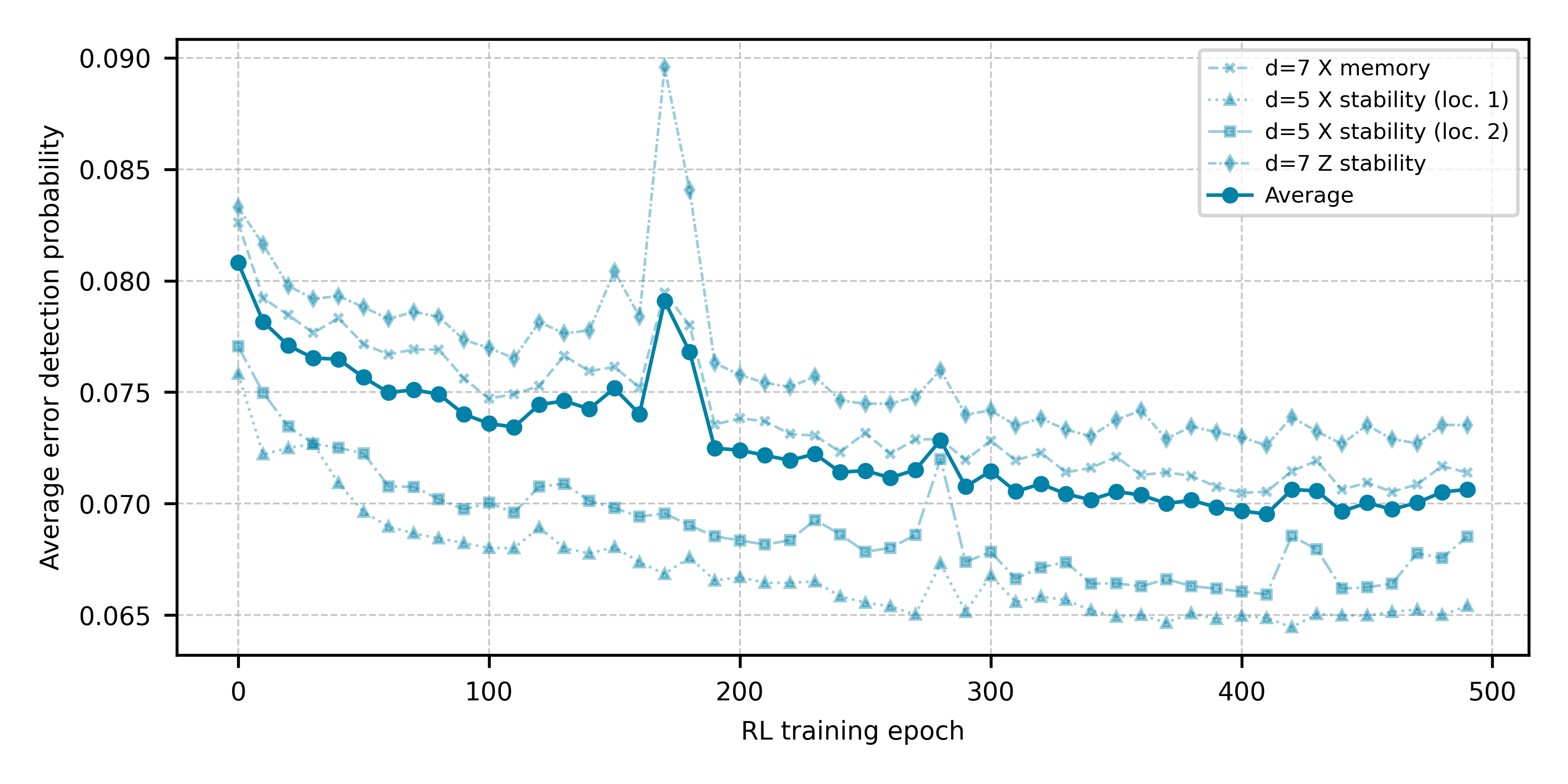}
    \caption{Evolution of average error detection probability during RL training on four QEC circuits.}
    \label{fig:rl_detections}
\end{figure}

We use reinforcement learning (RL) to fine-tune calibration parameters for QEC, as described in Ref.~\cite{sivak2026reinforcement}. In each RL epoch, we acquire QEC syndrome data from four circuits: distance-7 $X$-basis memory, distance-7 $Z$-basis stability, and distance-5 $X$-basis stability centered in two different locations. These circuits are selected to provide sufficient qubit coverage and ensure that control parameters are not overfit to any one circuit at the expense of losing performance in other circuits. We tune approximately $5,000$ system control parameters simultaneously, including those described in Ref.~\cite{sivak2026reinforcement} and additional new parameters. The error detection probability averaged across all detectors in these circuits decreases steadily during the course of training, as shown in Fig.~\ref{fig:rl_detections}. We save the QEC syndrome data obtained with the fine-tuned RL policy for further post-processing and the analysis shown in the main text.

\section{Extended Data} 
\label{sec:extended_data_xz_grids}

In the main text, logical error probabilities are averaged across processor subpatches and across the $X$ and $Z$ Pauli bases. Here we present the stability and surface code memory data resolved by individual subpatch and basis ($X$, $Z$, and their combined mean) across four hardware runs.

Table~\ref{tab:ext_shots_subsampling} summarizes the consolidated shot counts across the four runs. Surface code memory is measured in both the $X$ and $Z$ bases across $24$ cycle counts $c \in [2, 240]$ at all $15$ subpatches ($5 \times 10^4$ shots per run and cycle count). Stability is measured in both bases for $3 \times 3$, $4 \times 4$, and $5 \times 5$, and in the $Z$ basis only at $6 \times 6^*$ and $7 \times 7^*$ due to the processor boundary. Each stability run executes $5 \times 10^5$ shots of a $200$-cycle circuit; following Sec.~\ref{sec:supp_stability_subsampling}, we subsample $5$ evenly spaced, centered windows per shot within steady-state cycles $[22, 178)$ at each of $19$ window spans $c \in [2, 70]$.

\begin{figure*}[t!]
    \centering
    \includegraphics[width=\textwidth]{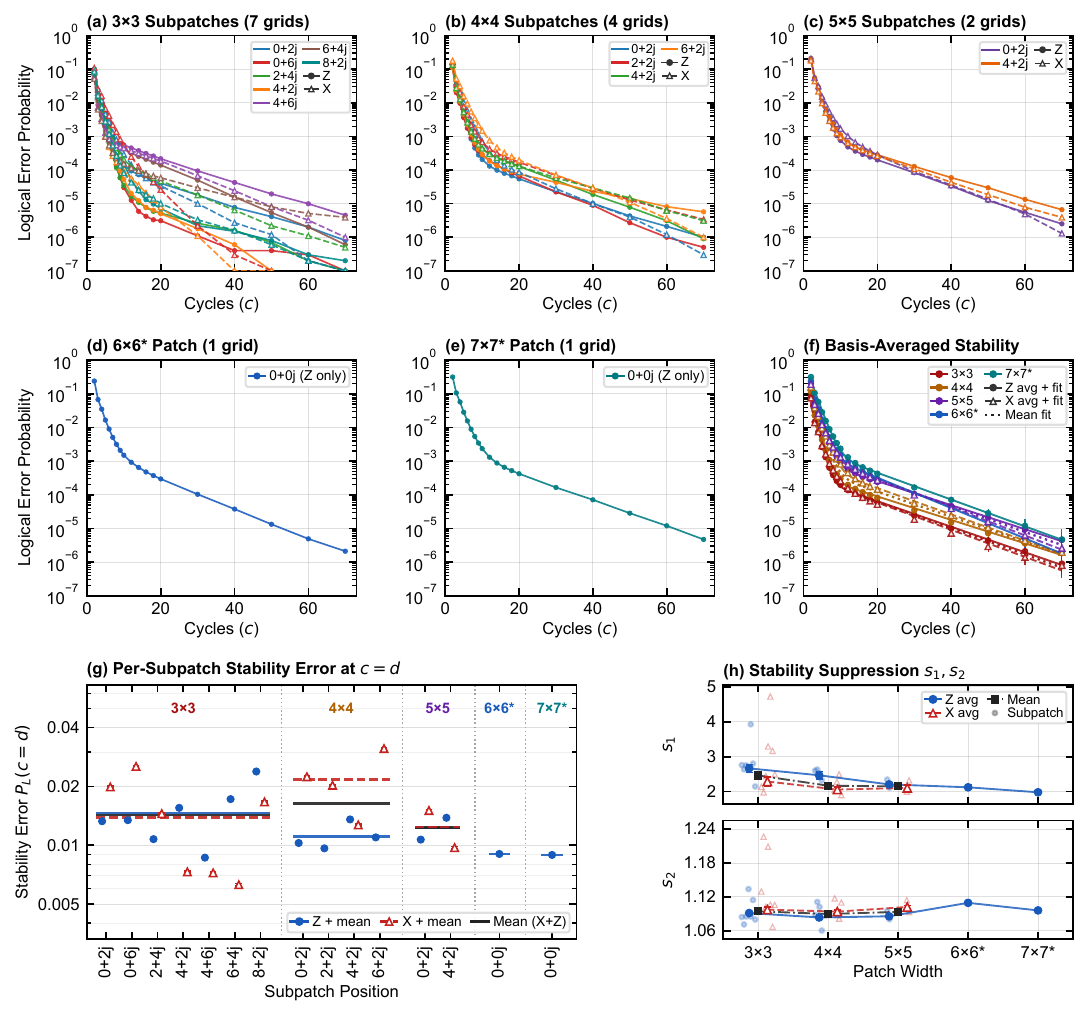}
    \caption{\textbf{Individual subpatch and $X$/$Z$ basis performance for the stability experiment.}
    \textbf{(a--e)} Logical error probability $P_L(c)$ versus cycles $c \in [2, 70]$ across individual subpatches for \textbf{(a)} $3 \times 3$ ($7$ grids), \textbf{(b)} $4 \times 4$ ($4$ grids), \textbf{(c)} $5 \times 5$ ($2$ grids), \textbf{(d)} $6 \times 6^*$ ($1$ grid), and \textbf{(e)} $7 \times 7^*$ ($1$ grid). Colors identify subpatches; solid curves and filled circles denote $Z$, and dashed curves and open triangles denote $X$ ($6 \times 6^*$ and $7 \times 7^*$ are $Z$ only).
    \textbf{(f)} Subpatch-averaged decay for $Z$ (circles, solid fits), $X$ (open triangles, dashed fits), and combined mean (dotted fits), fitted to $P_L(c) = a_1 s_1^{-c} + a_2 s_2^{-c}$.
    \textbf{(g)} Block error $P_L(c = d)$ across subpatches for $Z$ (blue circles) and $X$ (red triangles), with horizontal bars for the $Z$, $X$, and combined means.
    \textbf{(h)} Fast ($s_1$, top) and slow ($s_2$, bottom) suppression factors for individual subpatches (faint) and the $Z$, $X$, and combined means ($1\sigma$ error bars).}
    \label{fig:ext_stability}
\end{figure*}

Extended Data Fig.~\ref{fig:ext_stability} shows the stability logical error probability $P_L(c)$ for each subpatch and basis [panels (a--e)], the subpatch-averaged two-component fits $P_L(c) = a_1 s_1^{-c} + a_2 s_2^{-c}$ [panel (f)], the error $P_L(c = d)$ across subpatches [panel (g) and Table~\ref{tab:ext_xz_summary}], and the fast ($s_1$) and slow ($s_2$) suppression factors [panel (h)]. Every subpatch exhibits a rapid initial syndrome-suppression decay followed by a slower leakage-limited tail, with similar $X$ and $Z$ performance at odd patch widths and higher $X$-basis error at $4 \times 4$ due to its larger boundary footprint.
We note that the $X$ and $Z$ basis stability experiments have primary sensitivity to different sub-lattices of measure qubits.
As a result, measure qubit outliers (in leakage or Pauli error rates) can cause one basis to degrade while not significantly impacting the other. 

\begin{figure*}[t!]
    \centering
    \includegraphics[width=\textwidth]{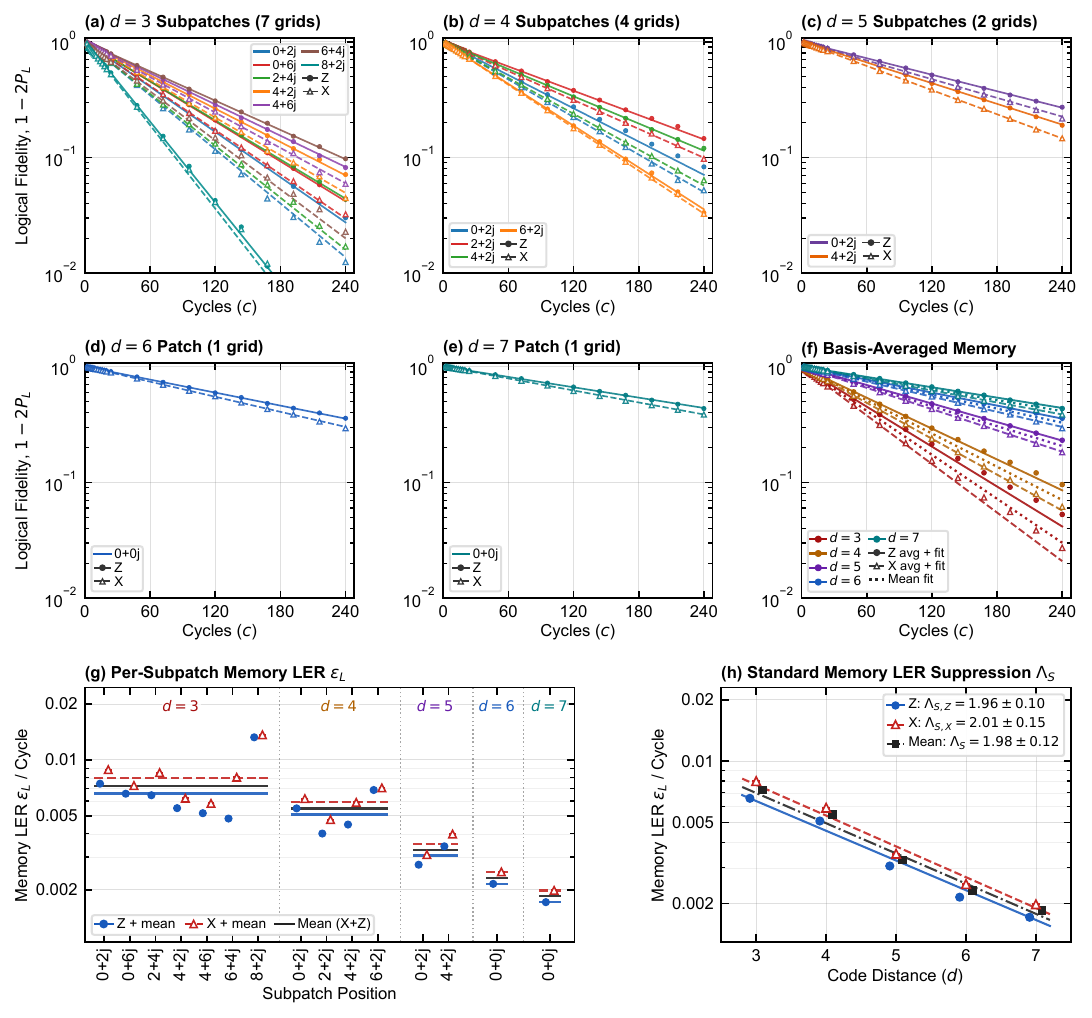}
    \caption{\textbf{Individual subpatch performance, $X$/$Z$ basis comparison, and standard LER suppression $\Lambda_S$ for surface code memory.}
    \textbf{(a--e)} Logical fidelity $1 - 2 P_L(c)$ versus cycles $c \in [2, 240]$ on a logarithmic axis across subpatches for \textbf{(a)} $d = 3$, \textbf{(b)} $d = 4$, \textbf{(c)} $d = 5$, \textbf{(d)} $d = 6$, and \textbf{(e)} $d = 7$. Colors match Extended Data Fig.~\ref{fig:ext_stability}, with filled circles and solid fits for $Z$, and open triangles and dashed fits for $X$.
    \textbf{(f)} Subpatch-averaged logical fidelity $1 - 2 P_L(c)$ for $Z$ (circles, solid fits), $X$ (open triangles, dashed fits), and combined mean (dotted fits).
    \textbf{(g)} Per-cycle logical error rate $\epsilon_L$ across subpatches for $Z$ (blue circles) and $X$ (red triangles), with $Z$, $X$, and combined mean bars.
    \textbf{(h)} Standard per-cycle LER suppression factor $\Lambda_S$ across $d \in \{3, 5, 7\}$ for $Z$, $X$, and combined mean (contrasted with the main-text block $\mu_S$ at $c = c^*$).}
    \label{fig:ext_memory}
\end{figure*}

Extended Data Fig.~\ref{fig:ext_memory} shows the corresponding surface code memory logical fidelity $1 - 2 P_L(c)$ across individual subpatches [panels (a--e)], the subpatch-averaged exponential fidelity fits [panel (f)], and the per-cycle logical error rates $\epsilon_L$ [panel (g) and Table~\ref{tab:ext_xz_summary}]. Across both experiments, changes in stability and memory performance across subpatches and bases are expected from qubit-to-qubit variation in both leakage rates and Pauli component error rates across the device. Finally, Extended Data Fig.~\ref{fig:ext_memory}(h) fits $\epsilon_L$ across odd distances $d \in \{3, 5, 7\}$ to extract the \emph{standard} per-cycle logical error rate (LER) suppression factor $\Lambda_S$ for the $Z$ basis, $X$ basis, and combined mean. We emphasize that $\Lambda_S$ shown here is the standard LER suppression factor characterizing the per-cycle rate $\epsilon_L$, whereas $\mu_S$ in the main text is the \emph{block} suppression factor evaluated from the optimal memory error probability $P^*_M = P_M(c^*)$ at the spacetime-optimal cycle count $c^*(d)$ (which includes data qubit state-preparation and measurement (SPAM) errors).

\begin{table*}[tb]
    \centering
    \small
    \setlength{\tabcolsep}{5pt}
    \caption{\textbf{Consolidated hardware shot counts and subsampled stability windows across four runs.} Totals are in units of $10^6$. Per subpatch and measured basis across the four runs, stability uses $2.0 \times 10^6$ raw $200$-cycle shots ($10.0 \times 10^6$ windows/span across $19$ spans $c \in [2, 70]$, $5$ windows/shot) and memory uses $0.20 \times 10^6$ shots/cycle across $24$ cycle counts $c \in [2, 240]$ in both $X$ and $Z$.}
    \label{tab:ext_shots_subsampling}
    \begin{tabular}{ccccccc}
        \hline\hline
        $d$ & Grids & \multicolumn{3}{c}{Stability ($\times 10^6$)} & \multicolumn{2}{c}{Memory ($X, Z$; $\times 10^6$)} \\
        \cline{3-5} \cline{6-7}
         & & Bases & Raw shots & Win./span & Shots/cyc. & Total ($24$ cyc.) \\
        \hline
        $3$ & $7$ & $X, Z$ & $28.0$ & $140.0$ & $2.80$ & $67.2$ \\
        $4$ & $4$ & $X, Z$ & $16.0$ & $80.0$ & $1.60$ & $38.4$ \\
        $5$ & $2$ & $X, Z$ & $8.0$ & $40.0$ & $0.80$ & $19.2$ \\
        $6^*$ & $1$ & $Z$ & $2.0$ & $10.0$ & $0.40$ & $9.6$ \\
        $7^*$ & $1$ & $Z$ & $2.0$ & $10.0$ & $0.40$ & $9.6$ \\
        \hline
        Total & $15$ & $28$ & $56.0$ & $280.0$ & $6.00$ & $144.0$ \\
        \hline\hline
    \end{tabular}
\end{table*}

\begin{table*}[tb]
    \centering
    \small
    \setlength{\tabcolsep}{4pt}
    \caption{\textbf{Basis-resolved stability and memory summary metrics.} Uncertainties denote $1\sigma$ standard errors. At $6 \times 6^*$ and $7 \times 7^*$, stability is measured in the $Z$ basis only.}
    \label{tab:ext_xz_summary}
    \begin{tabular}{cccccccc}
        \hline\hline
        $d$ & Grids & \multicolumn{3}{c}{Stability $P_L(c{=}d)$ ($\times 10^{-2}$)} & \multicolumn{3}{c}{Memory $\epsilon_L$/cyc. ($\times 10^{-3}$)} \\
        \cline{3-5} \cline{6-8}
         & & $Z$ & $X$ & Mean & $Z$ & $X$ & Mean \\
        \hline
        $3$ & $7$ & $1.47 \pm 0.01$ & $1.39 \pm 0.01$ & $1.43 \pm 0.01$ & $6.58 \pm 0.01$ & $7.99 \pm 0.01$ & $7.24 \pm 0.01$ \\
        $4$ & $4$ & $1.11 \pm 0.01$ & $2.17 \pm 0.01$ & $1.64 \pm 0.01$ & $5.09 \pm 0.01$ & $5.92 \pm 0.01$ & $5.49 \pm 0.01$ \\
        $5$ & $2$ & $1.23 \pm 0.01$ & $1.24 \pm 0.01$ & $1.23 \pm 0.01$ & $3.06 \pm 0.01$ & $3.51 \pm 0.01$ & $3.28 \pm 0.01$ \\
        $6^*$ & $1$ & $0.90 \pm 0.01$ & --- & $0.90 \pm 0.01$ & $2.15 \pm 0.01$ & $2.49 \pm 0.01$ & $2.32 \pm 0.01$ \\
        $7^*$ & $1$ & $0.89 \pm 0.01$ & --- & $0.89 \pm 0.01$ & $1.72 \pm 0.01$ & $1.99 \pm 0.01$ & $1.85 \pm 0.01$ \\
        \hline\hline
    \end{tabular}
\end{table*}

\end{document}